\documentclass[aip,jcp,reprint,noshowkeys,superscriptaddress]{revtex4-2}
\usepackage{graphicx,dcolumn,bm,microtype,multirow,amscd,amsmath,amssymb,amsfonts,physics,wrapfig,bbold,siunitx,xspace,ulem,txfonts,mathbbol}

\usepackage[version=4]{mhchem}

\usepackage[utf8]{inputenc}
\usepackage[T1]{fontenc}
\usepackage{feynmp-auto}
\usepackage{longtable}
\usepackage[dvipsnames]{xcolor}
\usepackage{mathtools}

\usepackage{hyperref}
\hypersetup{
    colorlinks,
    linkcolor={red!50!black},
    citecolor={red!70!black},
    urlcolor={red!80!black}
}

\usepackage{listings}
\definecolor{codegreen}{rgb}{0.58,0.4,0.2}
\definecolor{codegray}{rgb}{0.5,0.5,0.5}
\definecolor{codepurple}{rgb}{0.25,0.35,0.55}
\definecolor{codeblue}{rgb}{0.30,0.60,0.8}
\definecolor{backcolour}{rgb}{0.98,0.98,0.98}
\definecolor{mygray}{rgb}{0.5,0.5,0.5}

\definecolor{sqred}{rgb}{0.85,0.1,0.1}
\definecolor{sqgreen}{rgb}{0.25,0.65,0.15}
\definecolor{sqorange}{rgb}{0.90,0.50,0.15}
\definecolor{sqblue}{rgb}{0.10,0.3,0.60}

\lstdefinestyle{mystyle}{
    backgroundcolor=\color{backcolour},
    commentstyle=\color{codegreen},
    keywordstyle=\color{codeblue},
    numberstyle=\tiny\color{codegray},
    stringstyle=\color{codepurple},
    basicstyle=\ttfamily\footnotesize,
    breakatwhitespace=false,
    breaklines=true,
    captionpos=b,
    keepspaces=true,
    numbers=left,
    numbersep=5pt,
    numberstyle=\ttfamily\tiny\color{mygray},
    showspaces=false,
    showstringspaces=false,
    showtabs=false,
    tabsize=2
  }

  \newcolumntype{d}{D{.}{.}{-1}}

\usepackage{accents}

\newcommand{\Om}{\Omega}

\newcommand{\bC}{\boldsymbol{C}}

\newcommand{\bK}{\boldsymbol{K}}
\newcommand{\bQ}{\boldsymbol{Q}}

\newcommand{\bH}{\boldsymbol{H}}
\newcommand{\bO}{\boldsymbol{0}}
\newcommand{\bI}{\boldsymbol{1}}
\newcommand{\bU}{\boldsymbol{U}}
\newcommand{\bV}{\boldsymbol{V}}
\newcommand{\bff}{\boldsymbol{f}}

\newcommand{\bom}{\boldsymbol{\omega}}
\newcommand{\bSig}{\boldsymbol{\Sigma}}

\newcommand{\ii}{\text{i}}
\newcommand{\SupInf}{\textcolor{blue}{Supplementary Material}\xspace}

\newcommand{\LCPQ}{Laboratoire de Chimie et Physique Quantiques (UMR 5626), Universit\'e de Toulouse, CNRS, Toulouse, France}
\newcommand{\UnHam}{Department of Chemistry, University of Hamburg, 22761 Hamburg, Germany; The Hamburg Centre for Ultrafast Imaging (CUI), Hamburg 22761, Germany}

\begin{document}	

\title{An Algebraic-Diagrammatic Construction for Vertex Corrections to the $GW$ Self-Energy}

\author{Antoine \surname{Marie}}
	\email{amarie@irsamc.ups-tlse.fr}
	\affiliation{\LCPQ}

\author{Johannes \surname{T\"olle}}
        \email{johannes.toelle@uni-hamburg.de}
        \affiliation{\UnHam}
	
\author{Pierre-Fran\c{c}ois \surname{Loos}}
	\email{loos@irsamc.ups-tlse.fr}
	\affiliation{\LCPQ}
	
\begin{abstract}
  The $G3W2$ approximation --- the second-order self-energy beyond $GW$ --- is known to violate some fundamental analytic properties of the self-energy. 
  In particular, its lack of positive semi-definiteness leads to unphysical features such as negative spectral functions.
  In this work, we reformulate the $G3W2$ approximation within the algebraic-diagrammatic construction (ADC) framework.
  The resulting ADC-$G3W2$ formalism enforces the same analytic form as the exact self-energy, namely a sum-over-state representation, and, consequently, guarantees positive semi-definiteness.
  Starting from the $GW$ self-energy, we construct a hierarchy of ADC-based approximations of increasing sophistication, including ADC-2SOSEX, ADC(3)-$G3W2$, and a full ADC-$G3W2$ scheme. 
  These methods can be interpreted as nonperturbative resummations of vertex corrections to the self-energy, yielding Hermitian effective Hamiltonians whose diagonalization provides quasiparticle and satellite energies. 
  This establishes a formal bridge between many-body perturbation theory formulated in terms of the screened interaction $W$ and conventional ADC schemes based on the bare Coulomb interaction.
  The performance of these ADC-based approximations is gauged for valence ionization potentials and benchmarked against their parent method.
\bigskip
\begin{center}
	\boxed{\includegraphics[width=0.5\linewidth]{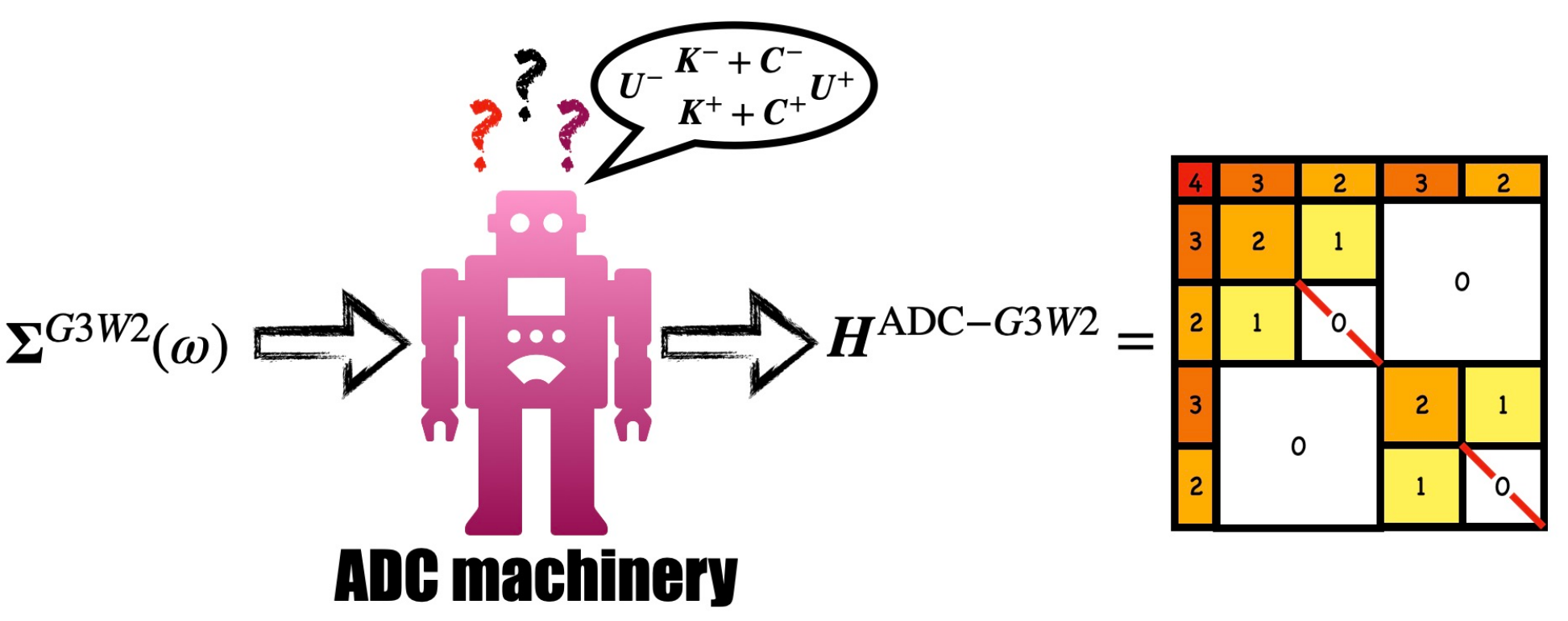}}
\end{center}
\bigskip
\end{abstract}

\maketitle

\section{Many-body Perturbation Theory}
\label{sec:mbpt}

Many-body perturbation theory (MBPT) \cite{Onida_2002,Martin_2016} has undergone rapid development in recent years. \cite{Reining_2017,Golze_2019,Marie_2024a,Blase_2018,Blase_2020} 
Although historically rooted in condensed matter physics, \cite{Strinati_1982a,Strinati_1982b,Hybertsen_1985a,Hybertsen_1986,Godby_1986,Godby_1987a} it has emerged as a viable and competitive framework for quantum chemical calculations on molecular systems. \cite{Shirley_1993,Rohlfing_2000,Stan_2006,Rostgaard_2010,Blase_2011b,Faber_2011,Ke_2011,Bruneval_2012,Bruneval_2013,vanSetten_2015}
In particular, MBPT-based approaches have become well-established tools for the computation of ionization potentials \cite{vanSetten_2015,Caruso_2016,Lewis_2019,Bruneval_2021,Monino_2023,Marie_2024b,vanSetten_2018,Golze_2018,Golze_2020,Mejia-Rodriguez_2021,Li_2022,Mukatayev_2023,Panades-Barrueta_2023} and optical excitations \cite{Grossman_2001,Tiago_2006,Rocca_2010,Blase_2011a,Baumeier_2012a,Duchemin_2012,Bruneval_2015,Leng_2016,Hirose_2015,Jacquemin_2017a,Jacquemin_2017b,Krause_2017,Liu_2020,McKeon_2022} at a computational cost that remains affordable \cite{Neuhauser_2014,Govoni_2015,Vlcek_2017,Wilhelm_2018,DelBen_2019,Forster_2020,Kaltak_2020,Forster_2021,Duchemin_2021,Wilhelm_2021,Forster_2022,Yu_2022,Tolle_2024a,Tolle_2024b} while achieving an accuracy comparable to that of wavefunction methods.

\begin{figure}
  \includegraphics[width=0.65\linewidth]{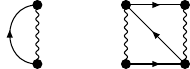}
  \caption{Diagrammatic representation of the $GW$ (left) and $G3W2$ (right) self-energies. 
  The solid and wiggly lines represent $G$ and $W$, respectively.}
  \label{fig:fig1}
\end{figure}

Within this context, the $GW$ approximation has proven especially successful. \cite{Hedin_1965,Aryasetiawan_1998,Reining_2017,Golze_2019,Marie_2024a}
In this approach, the one-body Green's function $G$ is obtained through the Dyson equation
\begin{equation}
  G^{-1} = G_0^{-1} - \Sigma
\end{equation}
(where $G_0$ is the Hartree Green's function) by approximating the self-energy $\Sigma$ as the product of $G$ and the dynamically screened Coulomb interaction $W$
\begin{equation}
  \Sigma^{GW} = \ii G W
\end{equation}
where $W = v + W_{\text{P}}$ is conveniently decomposed into its bare (static) Coulomb part $v$ and a (dynamic) polarizable contribution $W_{\text{P}}$.
This self-energy is represented diagrammatically in the left panel of Fig.~\ref{fig:fig1}.
The $GW$ approximation can be straightforwardly derived from Hedin's equations by retaining only the lowest-order contribution to the vertex function $\Gamma$. 
While this approximation captures a substantial fraction of dynamical correlation effects and provides a proper description of quasiparticle states in weakly-correlated systems, it remains inadequate for multireference systems, \cite{Ammar_2024,Wang_2026} strongly correlated materials,\cite{Verdozzi_1995,DiSabatino_2016,Tomczak_2017,Dvorak_2019a,Dvorak_2019b,DiSabatino_2022,DiSabatino_2023,Orlando_2023b} or multiparticle processes such as satellite excitations. \cite{Mejuto-Zaera_2022,Marie_2024b,Loos_2024}

Going beyond $GW$ is done through the inclusion of vertex corrections, either in the self-energy $\Sigma$ and/or in the polarizability $P$ (which is used to compute $W$).
Vertex corrections in the self-energy correspond to going beyond first order in the screened interaction $W$ and are commonly referred to as outer-vertex corrections, \cite{Romaniello_2012,Ren_2015,Pavlyukh_2020,Wang_2021a,Bruneval_2024,Bruneval_2025}
while vertex corrections in the polarizability improve the screening itself by incorporating interaction effects beyond the direct random-phase approximation (RPA) and are known as inner-vertex corrections. \cite{Northrup_1987,Shirley_1993,Shirley_1996,Bruneval_2005,Shishkin_2007b,Lewis_2019,Schmidt_2017,Bruneval_2021,Cunningham_2023} 
A growing body of evidence indicates that these two types of corrections should be treated on an equal footing: \cite{DelSol_1994,Schindlmayr_1998,Morris_2007,Romaniello_2009a,Gruneis_2014,Kutepov_2016,Kuwahara_2016,Hung_2017,Maggio_2017b,Vlcek_2017,Ma_2019,Tal_2021,Mejuto-Zaera_2022,Forster_2022a,Rohlfing_2023,Weng_2023,Wen_2024,Vacondio_2024,Forster_2024,Patterson_2024,Forster_2025} the apparent success of standard $GW$ often relies on a fortuitous cancellation of errors between $\Sigma$ and $P$. \cite{Lewis_2019,Mejuto-Zaera_2022,Bruneval_2024,Forster_2024,Forster_2025}
As a result, systematic improvements over $GW$ necessitate the consistent inclusion of both inner- and outer-vertex corrections, thereby providing a more balanced and physically sound description of electronic correlation.

However, it is often unclear how to systematically improve each of these quantities in a controlled and reliable manner.
For example, in the case of the self-energy, going beyond $GW$ typically amounts to adding higher-order terms in the perturbative expansion of $\Sigma$ with respect to $W$. \cite{Mejuto-Zaera_2022}
The leading correction beyond $GW$ is provided by the second-order self-energy, commonly referred to as the $G3W2$ approximation \cite{Kutepov_2021,Wang_2021a,Bruneval_2024,Bruneval_2025}
\begin{equation}
  \label{eq:g3w2_selfenergy}
  \Sigma^\text{$G3W2$} = \Sigma^{GW} + \ii^2 G W G W G
\end{equation}
This additional term is represented diagrammatically in the right panel of Fig.~\ref{fig:fig1}.
A natural approximation of this self-energy is obtained by discarding the computationally expensive $\ii^2 G W_{\text{P}} G W_{\text{P}} G$ contribution \cite{Bruneval_2024,Bruneval_2025}
\begin{equation}
  \label{eq:2sosex_selfenergy}
  \Sigma^\text{2SOSEX} = \Sigma^{GW} + \ii^2 G v G v G + \ii^2 G W_{\text{P}} G v G + \ii^2 G v G W_{\text{P}} G 
\end{equation}
This computationally more affordable approximation that retains the dominant second-order exchange and screening effects beyond $GW$ is referred to as the second-order screened exchange (2SOSEX) contribution.

Unfortunately, the convergence properties of such perturbative series are largely unknown, and the naive inclusion of higher-order contributions may lead to erratic or even deteriorated results.
Moreover, truncated perturbative expressions may violate fundamental properties of the self-energy.
For example, they are not guaranteed to have a sum-over-state representation, i.e., the same analytic form as the exact self-energy, \cite{Winter_1972} or they might not be positive semi-definite (psd).
This latter property of the self-energy is sufficient to ensure that the associated spectral function is positive. \cite{Cederbaum_1975}
These issues are not specific to perturbative expansions in terms of $W$, but were already recognized in early many-body perturbation theories formulated in terms of the bare Coulomb interaction $v$.\cite{Cederbaum_1975}

Stefanucci and co-workers have shown, using diagrammatic techniques, how to complete an approximate self-energy into a psd approximation by adding a finite number of diagrams. \cite{Stefanucci_2014,Uimonen_2015,Pavlyukh_2016}
Recently, this scheme has been applied to the 2SOSEX approximation, and the performance of the resulting 2SOSEX-psd self-energy has been gauged for ionization potentials of molecules. \cite{Bruneval_2025}
Note that the resulting psd self-energies obtained within this formalism do not necessarily admit a sum-over-state representation.

The algebraic-diagrammatic construction (ADC) goes one step further, as it is specifically designed to complete an approximate self-energy into a form that preserves the analytic structure of the exact self-energy.
This objective cannot, in general, be achieved by adding only a finite number of diagrams, as in the aforementioned scheme.
Instead, ADC relies on infinite resummations of diagrammatic contributions to enforce this property. 
This scheme was originally applied to the perturbation expansion of the self-energy in terms of the Coulomb interaction, and this led to the ADC($n$) family of approximations for the self-energy. \cite{Schirmer_1982,Schirmer_1983,Schirmer_2018}

In this work, we extend this idea by applying the ADC to the perturbative expansion of the self-energy in terms of $W$.
Within this perspective, ADC can be interpreted as a nonperturbative resummation of vertex corrections to the self-energy, yielding a hierarchy of ADC-based approximations that explicitly admit a sum-over-state representation and are, hence, positive semi-definite. \cite{Schirmer_2018}
This framework provides a route to improve upon $GW$ through the inclusion of vertex corrections while maintaining fundamental analytic properties of the self-energy.
A conceptually related, albeit distinct, strategy has recently been explored within the multichannel Dyson equation formalism \cite{Riva_2023,Riva_2024} to construct a screened formulation based on the static limit of $W$. \cite{Romaniello_2026}

Importantly, the present ADC treatment addresses only the outer-vertex part of the problem.
Identifying the appropriate inner-vertex corrections that should accompany this approach remains an open and essential question.
As will be discussed in the concluding section, a balanced treatment of both contributions is expected to improve accuracy and reduce the reliance on error cancellation, thereby paving the way toward a more complete and predictive MBPT framework.

\section{The algebraic-diagrammatic construction}
\label{sec:theoretical}

The $n$th-order ADC approximation for the self-energy, denoted ADC($n$), is defined as being complete through $n$th order in the Coulomb interaction and having a sum-over-states representation.
This latter property is enforced by adding and resumming a well-defined subset of higher-order contributions.
Solving the Dyson equation using an approximate ADC($n$) self-energy is referred to as Dyson-ADC($n$) in the literature. \cite{Schirmer_2018}

Formally, the ADC procedure starts from the exact sum-over-states representation of the self-energy.
The self-energy can be decomposed as
\begin{equation}
  \bSig(\omega) = \bSig(\infty) + \bSig^-(\omega) + \bSig^+(\omega)
\end{equation}
where $\bSig(\infty)$ is the static self-energy, while $\bSig^-(\omega)$ and $\bSig^+(\omega)$ are the lesser and greater, also known as hole and electron, branches of the dynamical part of the self-energy, respectively.
The branches of the dynamical self-energy have the following \textit{diagonal} spectral representation
\begin{equation}
  \bSig^\pm(\omega) = \qty(\bV^{\pm})^\dag \cdot \qty(\bom - \boldsymbol{\mathcal{E}}^{\pm} )^{-1} \cdot \bV^{\pm}
\end{equation}
where $\bV^{\pm}$ contains the so-called Dyson amplitudes, $\boldsymbol{\mathcal{E}}^{\pm}$ is a diagonal matrix gathering the poles of the self-energy in the lower/upper part of the complex plane, and $\bom = \omega \bI$.
Upon unitary transformations $\bQ^\pm$ such that
\begin{align}
  \bU^\pm & = \bQ^\pm \cdot \bV^\pm
  &
  \bK^\pm + \bC^\pm  & = \bQ^\pm \cdot \boldsymbol{\mathcal{E}}^\pm \cdot \qty(\bQ^\pm)^\dag
\end{align}
one can go from the diagonal spectral representation to the equivalent \textit{non-diagonal} ADC form 
\begin{equation}
  \label{eq:ADC_non-diag}
  \bSig^\pm(\omega) = \qty(\bU^\pm)^\dag \cdot \qty(\bom - \bK^\pm - \bC^\pm )^{-1} \cdot \bU^\pm
\end{equation}
where $\bK^\pm$ are diagonal matrices corresponding to the zeroth-order approximation of the diagonal matrices $\boldsymbol{\mathcal{E}}^\pm$.

So far, the discussion has been formulated in terms of the exact self-energy.
To construct the desired ADC($n$) scheme, the matrices $\bU^\pm$ and $\bC^\pm$ are expanded perturbatively as
\begin{subequations}
  \begin{align}
    \bU^\pm & = \bU^{\pm,(1)} + \bU^{\pm,(2)} + \cdots
    \\
    \bC^\pm  & = \bC^{\pm,(1)} + \bC^{\pm,(2)} + \cdots
  \end{align}
\end{subequations}
The corresponding perturbative expansion of the self-energy branches then reads
\begin{equation}
  \label{eq:ADC-series}
  \begin{split}
    \bSig^\pm(\omega) 
    & = \qty(\bU^\pm)^\dag \cdot \qty(\bom - \bK^\pm)^{-1} \cdot \sum_{k=0}^\infty \qty[ \bC^\pm \cdot \qty(\bom - \bK^\pm )^{-1} ]^k \cdot \bU^\pm
    \\
    & =\qty[\bU^{\pm,(1)}]^\dag \cdot \qty(\bom - \bK^\pm)^{-1} \cdot \bU^{\pm,(1)}
    \\
    & + \qty[\bU^{\pm,(2)}]^\dag \cdot \qty(\bom - \bK^\pm)^{-1} \cdot \bU^{\pm,(1)}
    \\
    & + \qty[\bU^{\pm,(1)}]^\dag \cdot \qty(\bom - \bK^\pm)^{-1} \cdot \bU^{\pm,(2)}
    \\
    & + \qty[\bU^{\pm,(1)}]^\dag \cdot \qty(\bom - \bK^\pm)^{-1} \cdot \bC^{\pm,(1)} \cdot \qty(\bom - \bK^\pm)^{-1} \cdot \bU^{\pm,(1)}
    \\
    & + \cdots
  \end{split}
\end{equation}
The perturbative expansions of blocks $\bU^\pm$ and $\bC^\pm$ are determined by explicitly identifying Eq.~\eqref{eq:ADC-series} with the corresponding perturbative expansion of the self-energy in powers of the bare Coulomb interaction. \cite{Schirmer_2018}
Once determined, these quantities can be inserted into Eq.~\eqref{eq:ADC_non-diag} to get the ADC($n$) approximate self-energy.

Upon substitution into the Dyson equation, and owing to its structure [see Eq.~\eqref{eq:ADC_non-diag}], the problem can be reformulated as a Hermitian eigenvalue problem. \cite{Schirmer_1982,Schirmer_1983,Schirmer_2018}
The eigenvalues of the resulting effective Hamiltonian correspond to the poles of the one-body Green's function, that is, the ionization potentials (IPs) and electron affinities (EAs) of the system.
Note that, for each type of charged excitations, this effective Hamiltonian provides both quasiparticle and satellite energies.

The main idea of this work is to apply this procedure to the $G3W2$ self-energy, thereby completing it into an approximation, referred to as ADC-$G3W2$, that retains the analytic structure of the exact self-energy.
The electron and hole branches of the $G3W2$ self-energy can both be written in terms of intermediate quantities as (see the \SupInf for a detailed derivation)
\begin{widetext}
  \begin{equation}
    \label{eq:G3W2-electron}
    \begin{split}
      \bSig^\pm(\omega) 
      & =\qty[ \bU^{\pm,(1)}_1 ]^{\dagger} \cdot \qty( \bom - \bK^{\pm}_1 )^{-1} \cdot \bU^{\pm,(1)}_1 
      \\
      & + \qty[ \bU^{\pm,(2)}_1 ]^{\dagger} \cdot \qty( \bom - \bK^{\pm}_1 )^{-1} \cdot \bU^{\pm,(1)}_1 
      + \qty[ \bU^{\pm,(1)}_1 ]^{\dagger} \cdot \qty( \bom - \bK^{\pm}_1 )^{-1} \cdot \bU^{\pm,(2)}_1 
      \\
      & + \qty[ \bU^{\pm,(1)}_1 ]^{\dagger} \cdot \qty( \bom - \bK^{\pm}_1 )^{-1} \cdot \bC^{\pm}_1 \cdot \qty( \bom- \bK^{\pm}_1 )^{-1} \cdot \bU^{\pm,(1)}_1 
      + \qty[ \bU^{\pm,(3)}_1 ]^{\dagger} \cdot \qty( \bom - \bK^{\pm}_1 )^{-1} \cdot \bU^{\pm,(1)}_1 
      + \qty[ \bU^{\pm,(1)}_1 ]^{\dagger} \cdot \qty( \bom - \bK^{\pm}_1 )^{-1} \cdot \bU^{\pm,(3)}_1 
      \\
      & + \qty[ \bU^{\pm,(1)}_2 + \bC^{\pm}_{2/1} \cdot \qty( \bom - \bK^{\pm}_1 )^{-1} \cdot \bU^{\pm,(1)}_1 ]^{\dagger} \cdot \qty( \bom - \bK^{\pm}_2 )^{-1} \cdot \qty[ \bU^{\pm,(1)}_2 + \bC^{\pm}_{2/1} \cdot \qty( \bom - \bK^{\pm}_1 )^{-1} \cdot \bU^{\pm,(1)}_1 ]
    \end{split}
  \end{equation}
\end{widetext}
The explicit expressions of all matrix elements for each branch will be detailed in Sec.~\ref{sec:ADC-GW}.
One can readily see that the $G3W2$ self-energy, when written in terms of these intermediate quantities, closely resembles the perturbative expansion of the self-energy in Eq.~\eqref{eq:ADC-series}.
Owing to this formulation, it becomes possible to identify the different contributions to the ADC blocks and to construct the corresponding ADC effective Hamiltonian based on the $G3W2$ self-energy.
For clarity, we deliberately refrain from assigning perturbative orders to the various terms, as these can be misleading in the present context and are not required for the construction of the ADC scheme.

\section{ADC matrix elements}
\label{sec:ADC-GW}

In the following, $\epsilon_p$ are the reference mean-field energies, $\Omega_{\nu}$ are the direct RPA excitation energies, $X_{ia,\nu}$ and $Y_{ai,\nu}$ the elements of the corresponding eigenvectors, $\braket{pq}{rs}$ are two-electron integrals in Dirac notations, and
\begin{equation}
  M_{pq,\nu} = \sqrt{2} \sum_{ia} \qty[ \braket{pa}{qi} X_{ia,\nu} +  \braket{pi}{qa} Y_{ai,\nu} ]
  \label{eq:M_pq}
\end{equation}
are the $GW$ effective integrals.
(We refer the reader to Ref.~\onlinecite{Marie_2024a} for more details.)
Here, the indices $p,q,r,s, \dots$ are used for arbitrary orbitals, $i,j,k,l$ label the occupied orbitals, $a,b,c,d$ denote the virtual orbitals, and $\mu, \nu, \dots $ are used for combined occupied-virtual indices.
All quantities are expressed in the spatial-orbital basis.

Retaining only the first term in $\bSig^+$ [see Eq.~\eqref{eq:G3W2-electron}], together with its counterpart in $\bSig^-$, recovers the standard $GW$ self-energy.
The ADC identification procedure naturally begins from this contribution, which defines the leading terms of the diagonal matrices $\bK^{-}$ and $\bK^{+}$,
\begin{subequations}
  \begin{align}
    \qty(\bK_1^-)_{i\nu,j\mu} & = \qty( \epsilon_{i} - \Omega_{\nu} ) \delta_{ij} \delta_{\nu\mu} 
    \\
    \qty(\bK_1^+)_{a\nu,b\mu} & = \qty( \epsilon_{a} + \Omega_{\nu} ) \delta_{ab} \delta_{\nu\mu}
  \end{align}
\end{subequations}
As usual, these contributions correspond to the inclusion of two-hole-one-particle (2h1p) and two-particle-one-hole (2p1h) configurations.
In this case, the energies of these configurations are ``dressed'', as they are constructed from RPA excitation energies.
The corresponding leading contributions to the coupling matrices are
\begin{subequations}
  \begin{align}
    \qty(\bU_1^{-,(1)})_{i\nu,q} &= M_{qi,\nu}^* 
    \\
    \qty(\bU_1^{+,(1)})_{a\nu,q} &= M_{aq,\nu}
  \end{align}
\end{subequations}
which are likewise dressed quantities (i.e., $GW$ effective integrals) in the present formalism. 
Once included in Eq.~\eqref{eq:ADC_non-diag}, solving the Dyson equation becomes equivalent to diagonalizing the following effective Hamiltonian~\cite{Schirmer_2018}
\begin{equation}
\label{eq:effective_ham_2h1p-2p1h}    
  \bH = 
  \begin{pmatrix}
    \bff      & \qty(\bU^{-}_1)^\dag  & \qty(\bU^{+}_1)^\dag  \\
    \bU^{-}_1 & \bK^{-}_1 + \bC^{-}_1 & \bO                   \\
    \bU^{+}_1 & \bO                   & \bK^{+}_1 + \bC^{+}_1 \\
  \end{pmatrix}
\end{equation}
with $\bC^{-}_1 = \bO$, $\bC^{+}_1 = \bO$, $\bU^{-}_1 = \bU^{-,(1)}_1$, and $\bU^{+}_1 = \bU^{+,(1)}_1$.
Here, $\bff$ represents the Fock matrix in the orbital basis.

The resulting ADC-$GW$ Hamiltonian corresponds to the standard $GW$ self-energy in its upfolded form. \cite{Bintrim_2021,Monino_2022,Quintero_2022,Monino_2023,Tolle_2022,Scott_2023,Marie_2023}
In other words, the associated ADC-$GW$ scheme is formally equivalent to the conventional $GW$ approximation.
In this case, the ADC procedure does not introduce any additional terms, which is expected since the $GW$ self-energy already admits a sum-over-state representation. \cite{Bruneval_2025}
This situation is closely analogous to the equivalence between ADC(2) and the second-order Green's function (GF2) self-energy. \cite{Schirmer_2018}

Going beyond the $GW$ approximation, one may apply the ADC procedure to the first three terms of $\bSig^\pm$.
These terms form the 2SOSEX-aug self-energy introduced in Ref.~\onlinecite{Bruneval_2025} and defined as the 2SOSEX self-energy [see Eq.~\eqref{eq:2sosex_selfenergy}] with one additional second-order exchange term of the form $\ii^2 GvGvG$.
This leads to additional contributions to the coupling blocks
\begin{subequations}
  \begin{align}
    \qty(\bU_1^{-,(2)})_{i\nu,q} &= \sum_{kc} \frac{M^*_{ck,\nu} \braket{ic}{kq}}{\epsilon_c - \epsilon_k + \Omega_{\nu}} + \sum_{kc} \frac{M^*_{kc,\nu} \braket{ik}{cq}}{\epsilon_c - \epsilon_k - \Omega_{\nu}}
    \\
    \qty(\bU_1^{+,(2)})_{a\nu,q} &= \sum_{kc} \frac{M_{ck,\nu} \braket{ak}{cq}}{\epsilon_c - \epsilon_k + \Omega_{\nu}} + \sum_{kc} \frac{M_{kc,\nu} \braket{ac}{kq}}{\epsilon_c - \epsilon_k - \Omega_{\nu}}
  \end{align}
\end{subequations}
that must be included in the effective Hamiltonian [see Eq.~\eqref{eq:effective_ham_2h1p-2p1h}].
The ADC scheme with all contributions determined so far is referred to as ADC-2SOSEX.
The composition of the various blocks entering the effective Hamiltonian in the different ADC-based schemes considered here is summarized in Table \ref{tab:blocks}.

\begin{table*}
\caption{Composition of the various blocks entering the different ADC-based schemes considered in this work.
See the main text for the explicit expression of each block.}
\label{tab:blocks}
\begin{ruledtabular}
\begin{tabular}{lccccc}
Blocks & Config.&	ADC-$GW$      &	ADC-2SOSEX      &	   ADC(3)-$G3W2$          &	ADC-$G3W2$	\\
\hline
$\bU^{-}_1$            & 2h1p & $\bU^{-,(1)}_1$	& $\bU^{-,(1)}_1 + \bU^{-,(2)}_1$ & $\bU^{-,(1)}_1 + \bU^{-,(2)}_1$ & $\bU^{-,(1)}_1 + \bU^{-,(2)}_1 + \bU^{-,(3)}_1$ \\
$\bU^{+}_1$            & 2p1h & $\bU^{+,(1)}_1$	& $\bU^{+,(1)}_1 + \bU^{+,(2)}_1$ & $\bU^{+,(1)}_1 + \bU^{+,(2)}_1$ & $\bU^{+,(1)}_1 + \bU^{+,(2)}_1 + \bU^{+,(3)}_1$ \\
$\bK^{-}_1+\bC^{-}_1$  & 2h1p &	$\bK^{-}_1$     & $\bK^{-}_1$                     & $\bK^{-}_1 + \bC^{-,(1)}_1$     & $\bK^{-}_1 + \bC^{-,(1)}_1$                     \\
$\bK^{+}_1+\bC^{+}_1$  & 2p1h &	$\bK^{+}_1$     & $\bK^{+}_1$                     & $\bK^{+}_1 + \bC^{+,(1)}_1$     & $\bK^{+}_1 + \bC^{+,(1)}_1$                     \\
\hline
$\bU^{-}_2$            & 3h2p & $\bO$ & $\bO$ & $\bO$ & $\bU^{-,(1)}_2$     \\
$\bU^{+}_2$            & 3p2h & $\bO$ & $\bO$ & $\bO$ & $\bU^{+,(1)}_2$     \\
$\bK^{-}_2 +\bC^{-}_2$ & 3h2p & $\bO$ & $\bO$ & $\bO$ & $\bK^{-}_2$         \\
$\bK^{+}_2 +\bC^{+}_2$ & 3p2h & $\bO$ & $\bO$ & $\bO$ & $\bK^{+}_2$         \\
\hline
$\bC^{-}_{1/2}$        & 2h1p/3h2p & $\bO$ & $\bO$ & $\bO$ & $\bC^{-,(1)}_{1/2}$ \\
$\bC^{+}_{1/2}$        & 2p1h/3p2h & $\bO$ & $\bO$ & $\bO$ & $\bC^{+,(1)}_{1/2}$
\end{tabular}
\end{ruledtabular}
\end{table*}

When including the corresponding contributions in Eq.~\eqref{eq:ADC_non-diag}, one can see that the ADC-2SOSEX self-energy has two additional terms,
$\qty[ \bU^{-,(2)}_1 ]^{\dagger} \cdot \qty( \bom - \bK^{-}_1 )^{-1} \cdot \bU^{-,(2)}_1$ and
$\qty[ \bU^{+,(2)}_1 ]^{\dagger} \cdot \qty( \bom - \bK^{+}_1 )^{-1} \cdot \bU^{+,(2)}_1$,
compared to the 2SOSEX-aug self-energy.
In this case, the ADC procedure generates only a finite number of additional contributions because there are no contributions in the $\bC^\pm$ blocks so far [see Eq.~\eqref{eq:ADC-series}].
Note that ADC-2SOSEX is exactly equivalent to the $GW$+2SOSEX-psd self-energy reported in Ref.~\onlinecite{Bruneval_2025} and derived using the procedure of Refs.~\onlinecite{Stefanucci_2014,Uimonen_2015,Pavlyukh_2016}.

Proceeding towards the full $G3W2$ self-energy, the next contributions to be included correspond to the fourth term in Eq.~\eqref{eq:G3W2-electron}, which determines the first contribution to the diagonal blocks
\begin{subequations}
\begin{align}
  \qty(\bC^{-,(1)}_1)_{i\nu,j\mu} &= \frac{1}{2} \sum_{c} \frac{M_{ic,\mu} M^*_{jc,\nu}}{\epsilon_i - \epsilon_c + \Omega_{\mu}} + \frac{1}{2} \sum_{c} \frac{M_{ic,\mu} M^*_{jc,\nu}}{\epsilon_j - \epsilon_c + \Omega_{\nu}}
  \\
  \qty(\bC^{+,(1)}_1)_{a\nu,b\mu} &= \frac{1}{2} \sum_{k} \frac{M^*_{ka,\mu} M_{kb,\nu}}{\epsilon_a - \epsilon_k - \Omega_{\mu}} + \frac{1}{2} \sum_{k} \frac{M^*_{ka,\mu} M_{kb,\nu}}{\epsilon_b - \epsilon_k - \Omega_{\nu}}
\end{align}
\end{subequations}
The ADC scheme incorporating all contributions identified thus far will be referred to as ADC(3)-$G3W2$ (see Table \ref{tab:blocks}).
At this stage, the inclusion of $\bC$ in the resolvent $(\bom - \bK - \bC)^{-1}$ effectively generates an infinite resummation of diagrams, reflecting the nonperturbative character of the ADC construction beyond zeroth order.

Finally, applying the ADC procedure to the full $G3W2$ self-energy [see Eq.~\eqref{eq:G3W2-electron}] yields a new approximation to the self-energy, which we denote ADC-$G3W2$.
The remaining $G3W2$ self-energy terms not yet accounted for provide a third contribution to the coupling blocks
\begin{subequations}
  \begin{align}
    \begin{split}
      \qty(\bU_1^{-,(3)})_{i\nu,q}
      & = \frac{1}{2} \sum_{kc\mu} \frac{M_{ic,\mu} M^*_{kc,\nu} M^*_{qk,\mu}}{(\epsilon_c - \epsilon_k - \Omega_{\nu})(\epsilon_c - \epsilon_i - \Omega_{\mu})} 
      \\
      & - \sum_{kc\mu} \frac{M^*_{ci,\mu} M^*_{kc,\nu} M_{kq,\mu}}{(\epsilon_c - \epsilon_k - \Omega_{\nu})(\epsilon_c - \epsilon_i + \Omega_{\mu})} 
      \\
      & - \sum_{kc\mu} \frac{M^*_{ki,\mu} M^*_{ck,\nu} M_{cq,\mu}}{(\epsilon_c - \epsilon_i + \Omega_{\nu} + \Omega_{\mu})(\epsilon_c - \epsilon_k + \Omega_{\nu})} 
      \\
      & + \sum_{cd\mu} \frac{M^*_{di,\mu} M^*_{cd,\nu} M_{cq,\mu}}{(\epsilon_c - \epsilon_i + \Omega_{\nu} + \Omega_{\mu})(\epsilon_d - \epsilon_i + \Omega_{\mu})}
    \end{split}
    \\
    \begin{split}
      \qty(\bU_1^{+,(3)})_{a\nu,q} 
      & = \frac{1}{2} \sum_{kc\mu} \frac{M^*_{ka,\mu} M_{kc,\nu} M_{cq,\mu}}{(\epsilon_c - \epsilon_k - \Omega_{\nu})(\epsilon_a - \epsilon_k - \Omega_{\mu})} 
      \\
      & - \sum_{kc\mu} \frac{M_{ak,\mu} M_{kc,\nu} M^*_{qc,\mu}}{(\epsilon_c - \epsilon_k - \Omega_{\nu})(\epsilon_a - \epsilon_k + \Omega_{\mu})} 
      \\
      & - \sum_{kc\mu} \frac{M_{ac,\mu} M_{ck,\nu} M^*_{qk,\mu} }{(\epsilon_a - \epsilon_k + \Omega_{\nu} + \Omega_{\mu})(\epsilon_c - \epsilon_k + \Omega_{\nu})} 
      \\
      & + \sum_{kl\mu} \frac{M_{al,\mu} M_{lk,\nu} M^*_{qk,\mu}}{(\epsilon_a - \epsilon_k + \Omega_{\nu} + \Omega_{\mu})(\epsilon_a - \epsilon_l + \Omega_{\mu})}
    \end{split}
  \end{align}
\end{subequations}
They also introduce additional diagonal blocks corresponding to the inclusion of three-hole-two-particle (3h2p) and three-particle-two-hole (3p2h) configurations
\begin{subequations}
\begin{align}
  \qty(\bK_2^-)_{i\nu\mu,j\lambda\sigma} & = \qty( \epsilon_{i} - \Omega_{\nu} - \Omega_{\mu} ) \delta_{ij} \delta_{\nu\lambda} \delta_{\mu\sigma} 
  \\
  \qty(\bK_2^+)_{a\nu\mu,b\lambda\sigma} & = \qty( \epsilon_{a} + \Omega_{\nu} + \Omega_{\mu} ) \delta_{ab} \delta_{\nu\lambda} \delta_{\mu\sigma}
\end{align}
\end{subequations}
together with the corresponding coupling blocks
\begin{subequations}
\begin{align}
  \qty(\bU_2^{-,(1)})_{i\nu\mu,q} &= - \sum_{c} \frac{M^*_{ci,\mu} M^*_{qc,\nu}}{\epsilon_c - \epsilon_i + \Omega_{\mu}} 
  \\
  \qty(\bU_2^{+,(1)})_{a\nu\mu,q} &= + \sum_{k} \frac{M_{ak,\mu} M_{kq,\nu}}{\epsilon_a - \epsilon_k + \Omega_{\mu}}
\end{align}
\end{subequations}
Finally, couplings arise between the 2h1p and 3h2p configurations, as well as between the 2p1h and 3p2h configurations
\begin{subequations}
\begin{align}                               
  \qty(\bC^{-,(1)}_{2/1})_{i\nu\mu,j\lambda} &= M^*_{ji,\mu} \delta_{\nu\lambda}
  \\
  \qty(\bC^{+,(1)}_{2/1})_{a\nu\mu,b\lambda} &= M_{ab,\mu} \delta_{\nu\lambda}
\end{align}
\end{subequations}

As a result, the effective Hamiltonian must be expanded to include the 3h2p and 3p2h configurations with respect to Eq.~\eqref{eq:effective_ham_2h1p-2p1h}, leading to
\begin{equation}
    \bH = 
    \begin{pmatrix}
      \bff      & \qty(\bU^{-}_1)^\dag  & \qty(\bU^{-}_2)^\dag     & \qty(\bU^{+}_1)^\dag  & \qty(\bU^{+}_2)^\dag     \\
      \bU^{-}_1 & \bK^{-}_1 + \bC^{-}_1 & \qty(\bC^{-}_{2/1})^\dag & \bO                   & \bO                      \\
      \bU^{-}_2 & \bC^{-}_{2/1}         & \bK^{-}_2 + \bC^{-}_2    & \bO                   & \bO                      \\
      \bU^{+}_1 & \bO                   & \bO                      & \bK^{+}_1 + \bC^{+}_1 & \qty(\bC^{+}_{2/1})^\dag \\
      \bU^{+}_2 & \bO                   & \bO                      & \bC^{+}_{2/1}         & \bK^{+}_2 + \bC^{+}_2    \\
    \end{pmatrix} 
\end{equation}
Starting from Eq.~\eqref{eq:G3W2-electron}, ADC-$G3W2$ thus performs an infinite resummation of diagrams, yielding an approximation that admits a sum-over-state representation.

As can be seen from the above equation, the effective Hamiltonian associated with this ADC-based resummation is Hermitian. 
This property follows from the particular expression of the self-energy [see Eq.\eqref{eq:G3W2-electron}] used to derive the various blocks. 
Specifically, this expression was obtained following the procedure introduced in Ref.~\onlinecite{Bruneval_2025}, where the authors showed that decomposing dressed poles of the form $\omega - \epsilon_p \pm \Om_\nu$ together with bare poles (i.e., involving three single-particle energies) in the $G3W2$ self-energy yields a self-energy expressed solely in terms of dressed poles.
The ADC procedure could, in principle, have been applied directly to the $G3W2$ self-energy prior to this decomposition. 
However, the resulting resummation would then be associated with a non-Hermitian effective Hamiltonian. 
Although the decomposition strategy can, in principle, be extended to any vertex correction to the self-energy, there is no guarantee that a form yielding a Hermitian effective Hamiltonian can always be found.
Finally, it is worth noting that the resummation performed prior to pole decomposition can be described using the standard Goldstone diagrammatic rules. 
By contrast, the terms appearing in Eq.~\eqref{eq:G3W2-electron} after decomposition cannot be represented within this framework.


\section{Computational details}
\label{sec:comp_det}

Multiple implementations of the various schemes have been benchmarked using two complementary approaches: either full diagonalization of the resulting Hermitian matrices, performed with the \textsc{quack} software, \cite{QuAcK} or a root-following Davidson algorithm implemented in a Python code that heavily relies on functionalities provided by PySCF. \cite{Sun_2018,Sun_2020}
The computational scaling of the Davidson-based implementation for the determination of a single root is reported in Table~\ref{tab:Scaling}, together with an example of a contraction/intermediate illustrating the associated computational complexity.
Note that the diagonalization of the RPA eigenvalue problem, which scales as $\order*{K^6}$ ($K$ being the number of basis functions), and the integral transformation, which scales as $\order*{K^5}$, required to construct $M_{pq,\nu}$ are not included in these estimates, as they are common to all methods.

In all calculations, Hartree-Fock orbitals and energies are used as a starting point, the broadening parameter $\eta$ is set to zero, and the Davidson solver used to compute the quasiparticle energies is converged to a threshold of \SI{d-8}{\hartree}.
To avoid numerical instabilities associated with small energy denominators, we employ an energy-dependent regularization scheme with the flow parameter fixed at $s = \num{d6}$. \cite{Marie_2023,Marie_2025b}
This large value of $s$ corresponds to an extremely weak regularization, effectively leading to unregularized results.
The 2SOSEX and $G3W2$ calculations have been performed with \textsc{quack} and the implementation has been verified against \textsc{molgw}. \cite{Bruneval_2016}
The corresponding quasiparticle energies have been obtained by solving the frequency-dependent quasiparticle equation within the diagonal approximation, i.e., for each orbital independently.
Moreover, we have implemented the Dyson version of ADC(2) and ADC(3) in their spin-adapted form. \cite{Schirmer_1983,vonNiessen_1984}
Non-Dyson ADC(2) and ADC(3) \cite{Schirmer_1998,Trofimov_2002,Schirmer_1991,Mertins_1996a,Mertins_1996b} calculations have been performed with PySCF. \cite{Sun_2018,Sun_2020}

\begin{table}
\caption{Computational scaling of the various ADC schemes presented in this work, together with an example of contraction/intermediate illustrating the origin of their overall scaling. 
The diagonalization of the RPA eigenvalue problem [$\order*{K^6}$] and the integral transformation [$\order*{K^5}$] required to construct $M_{pq,\nu}$ [see Eq.~\eqref{eq:M_pq}] are common to all methods and are therefore not included in these estimates. 
Here, $K$ denotes the number of basis functions and $r^{(n)}_{b\nu}$ an element of the $n$th excitation vector.}
\label{tab:Scaling}
\begin{ruledtabular}
\begin{tabular}{lcc}
Method & Scaling &	Contraction / Intermediate\\
\hline
ADC-$GW$      &	$\order*{K^4}$ & $\sum_{b\nu} M_{ab,\nu} r^{(n)}_{b\nu}$ \\
ADC-2SOSEX    &	$\order*{K^5}$ & $\sum_{\nu} \frac{M_{ck,\nu}}{\epsilon_c - \epsilon_k + \Omega_{\nu}} r^{(n)}_{b\nu}$ \\
ADC(3)-$G3W2$ &	$\order*{K^5}$ & $\sum_{\nu} \frac{M_{ck,\nu}}{\epsilon_c - \epsilon_k + \Omega_{\nu}} r^{(n)}_{b\nu}$ \\    
ADC-$G3W2$	  &	$\order*{K^7}$ &  $\frac{M^*_{di,\mu}}{ (\epsilon_c - \epsilon_i + \Omega_{\nu} + \Omega_{\mu}) (\epsilon_d - \epsilon_i + \Omega_{\mu})}$\\
\end{tabular}
\end{ruledtabular}
\end{table}

\begin{figure*}
	\includegraphics[width=0.3\linewidth]{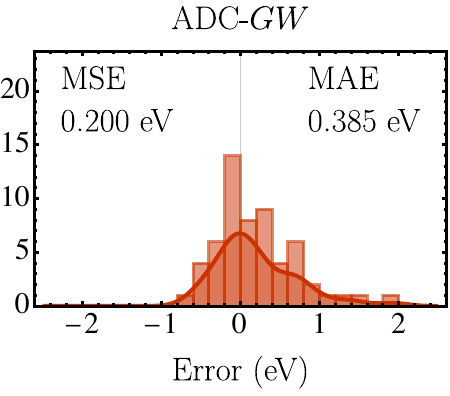}
	\hspace{0.02\linewidth}
	\includegraphics[width=0.3\linewidth]{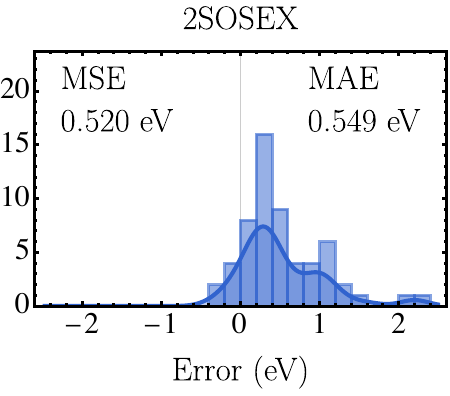}
	\hspace{0.02\linewidth}
	\includegraphics[width=0.3\linewidth]{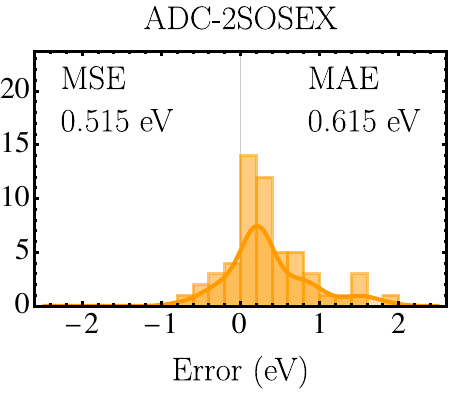}
	\\
	\vspace{0.02\textheight}
	\includegraphics[width=0.3\linewidth]{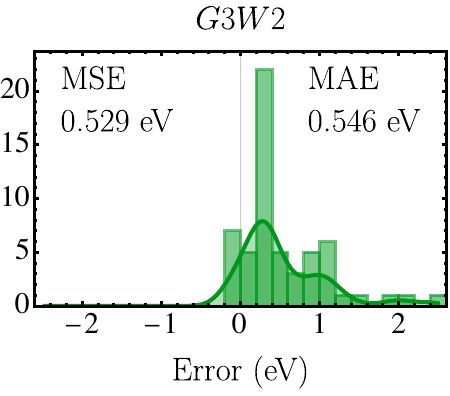}
	\includegraphics[width=0.3\linewidth]{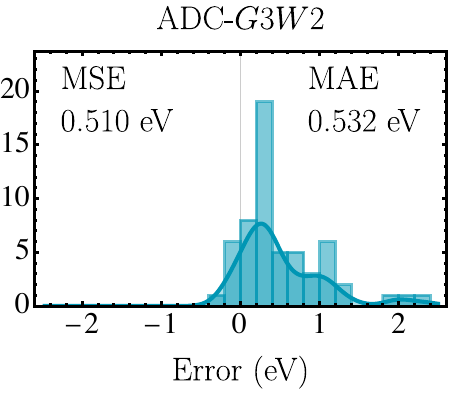}
	\caption{Histogram of the errors (with respect to the TBEs) for the inner- and outer-valence IPs computed with various self-energy schemes in the aug-cc-pVDZ basis set.
	All calculations are performed in the diagonal approximation.}
	\label{fig:diag}
\end{figure*}

\section{Results}
\label{sec:results}


Here, we consider the set of inner- and outer-valence IPs designed in Ref.~\onlinecite{Marie_2024b}, which is composed of 58 valence IPs computed in small molecular systems.
The molecular geometries have been extracted from the same study.
The reference IPs, the so-called theoretical best estimates (TBEs), are computed at the full configuration interaction (FCI) level and are thus highly accurate.
All IPs are computed in the aug-cc-pVDZ basis set.

First, we compute this set of IPs within the diagonal approximation using the following methods: ADC-$GW$ (equivalent to the one-shot $G_0W_0$ scheme), 2SOSEX, ADC-2SOSEX (equivalent to 2SOSEX-psd\cite{Bruneval_2025}), $G3W2$, and ADC-$G3W2$.
The diagonal approximation in ADC schemes corresponds to restricting the Fock matrix to a single orbital energy and diagonalizing it for each orbital.
We then consider several ADC schemes, namely ADC-$GW$, ADC-2SOSEX, ADC(3)-$G3W2$, and ADC-$G3W2$, beyond the diagonal approximation, i.e., the effective Hamiltonian is diagonalized once, and all quasiparticle energies are obtained from a single calculation.
All the raw data can be found in the \SupInf.
In addition, we evaluate the same set of IPs at the ADC(2) and ADC(3) levels within the Dyson formalism. 
In this framework, the hole and electron sectors are coupled, as in the ADC schemes developed in the present study. 
For completeness, the \SupInf also reports the corresponding non-Dyson ADC(2) and ADC(3) results, in which the two sectors are decoupled. 
These results show that, while Dyson and non-Dyson ADC(2) yield similar IPs, the differences become significantly more pronounced at the ADC(3) level (see below). \cite{Trofimov_2005,Dreuw_2023,Banerjee_2023}

Figure~\ref{fig:diag} reports the error distribution (with respect to the TBEs) for the various self-energies within the diagonal approximation. 
The $GW$ approximation yields a MAE of \SI{0.385}{\eV} and a MSE of \SI{0.200}{\eV}, in agreement with previous reports for this set of IPs. \cite{Marie_2024b,Loos_2026}
The 2SOSEX and ADC-2SOSEX approximations yield similar MSEs (\SI{0.520}{\eV} and \SI{0.515}{\eV}, respectively), while the MAE of ADC-2SOSEX (\SI{0.615}{\eV}) is slightly larger than that of its 2SOSEX counterpart (\SI{0.559}{\eV}). 
This trend contrasts with the findings of Ref.~\onlinecite{Bruneval_2025}, where Bruneval and coworkers reported that, using the same set of Hartree-Fock starting orbitals, the 2SOSEX-psd self-energy provides more accurate results than 2SOSEX for the $GW100$ benchmark set (MAEs of \SI{0.29}{\eV} and \SI{0.49}{\eV}, respectively). 
It is worth noting that $GW100$ contains only principal IPs, whereas the present set additionally includes inner-valence excitations, which are significantly more challenging to describe, even at the $GW$ level. 
Finally, $G3W2$ yields a MAE of \SI{0.546}{\eV} and a MSE of \SI{0.529}{\eV}, in line with results obtained for the $GW100$ benchmark, where a MAE of \SI{0.51}{\eV} was reported. \cite{Bruneval_2024}
Within the diagonal approximation, ADC-G3W2 yields a MAE of \SI{0.532}{\eV} and a MSE of \SI{0.510}{\eV}, slightly improving these statistical indicators without substantially altering the overall picture.

\begin{figure*}
	\includegraphics[width=0.3\linewidth]{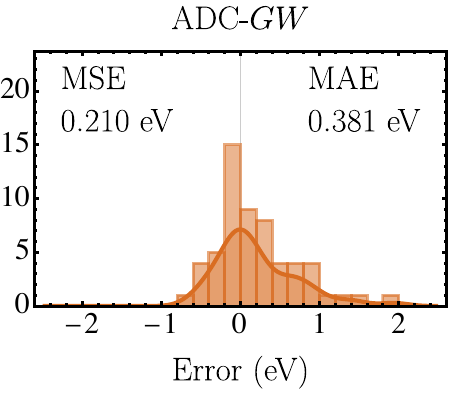}
	\hspace{0.02\linewidth}
	\includegraphics[width=0.3\linewidth]{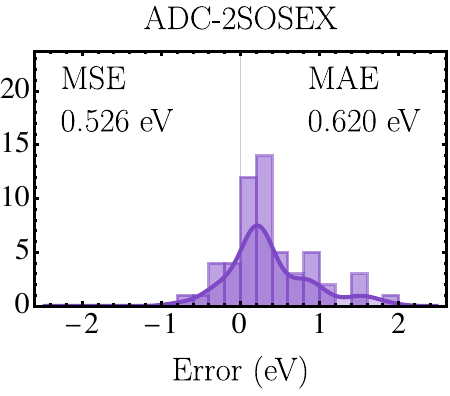}
	\hspace{0.02\linewidth}
	\includegraphics[width=0.3\linewidth]{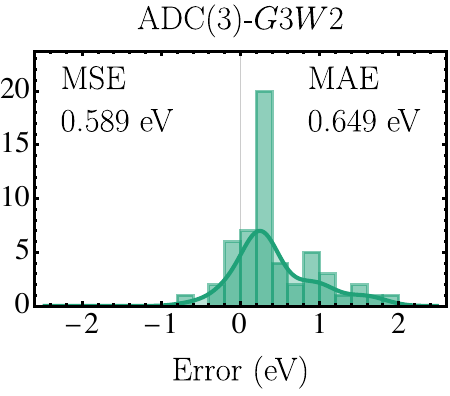}
	\\
	\vspace{0.02\textheight}
	\includegraphics[width=0.3\linewidth]{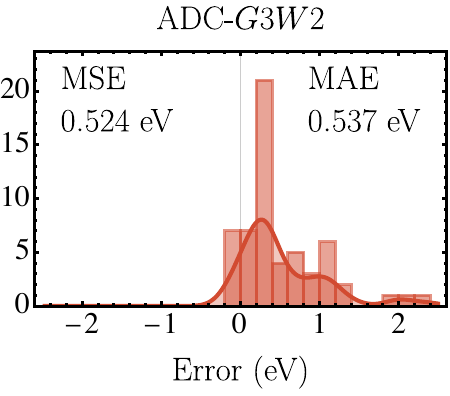}
	\hspace{0.02\linewidth}
	\includegraphics[width=0.3\linewidth]{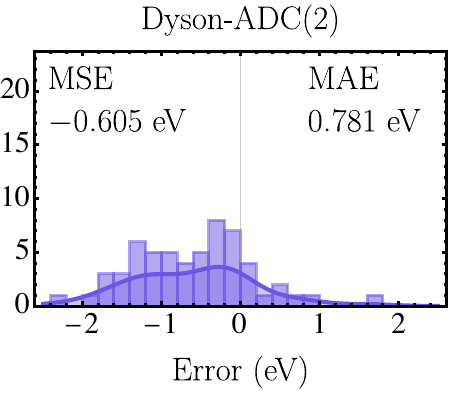}
	\hspace{0.02\linewidth}
	\includegraphics[width=0.3\linewidth]{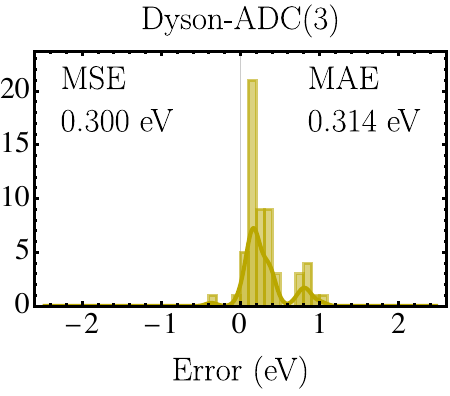}
	\caption{Histogram of the errors (with respect to the TBEs) for the inner- and outer-valence IPs computed with various ADC schemes in the aug-cc-pVDZ basis set.
	All calculations are performed without diagonal approximation.}
	\label{fig:ADC}
\end{figure*}

In Fig.~\ref{fig:ADC}, we report the error distribution (with respect to the TBEs) for the various ADC schemes. 
Among these, ADC-$GW$ yields the smallest MAE of \SI{0.381}{\eV}. 
This MAE increases to \SI{0.620}{\eV} for ADC-2SOSEX and further to \SI{0.649}{\eV} for ADC(3)-$G3W2$, before decreasing to \SI{0.537}{\eV} for ADC-$G3W2$. 
A similar trend is observed for the MSEs, which are all positive and of comparable magnitude to the corresponding MAEs (see Fig.\ref{fig:ADC}). 
For comparison, ADC(2) and ADC(3) yield MAEs of \SI{0.781}{\eV} and \SI{0.314}{\eV}, respectively, with MSEs of opposite sign [\SI{-0.605}{\eV} for ADC(2) and \SI{0.300}{\eV} for ADC(3)]. 
For the corresponding non-Dyson variants, the MAEs are \SI{0.851}{\eV} for ADC(2) and \SI{0.404}{\eV} for ADC(3), while the MSEs amount to \SI{-0.683}{\eV} and \SI{0.293}{\eV}, respectively (see the \SupInf). 
As mentioned earlier, decoupling the hole and electron sectors has little effect on the IPs at the ADC(2) level, whereas it noticeably increases the MAE at the ADC(3) level.
As anticipated, ADC-$G3W2$ is found to be less accurate than the conventional ADC-$GW$ scheme. 
Nevertheless, the inclusion of inner-vertex corrections is expected to improve the performance of ADC-$G3W2$, while the accuracy of ADC-$GW$ is known to deteriorate as the description of screening is refined. \cite{Lewis_2019,Forster_2024,Forster_2025}
We hope to report on such inner-vertex corrections within ADC-$G3W2$ in future work.

\section{Conclusion}
\label{sec:conclusion}

The present work may be viewed as an ADC-based treatment of outer-vertex corrections to the self-energy, thereby providing a beyond-$GW$ self-energy that retains the correct sum-over-states representation.
A natural and important next step is therefore to identify the corresponding inner-vertex corrections that should accompany this approach, in order to restore consistency between the self-energy and the polarizability. 
Such a balanced treatment is expected to improve the accuracy of the method. 

From a practical perspective, non-Dyson variants of the ADC-$G3W2$ schemes appear attractive, as they avoid the need for a root-following Davidson algorithm. \cite{Schirmer_1998,Trofimov_2002,Schirmer_1991,Mertins_1996a,Mertins_1996b}
While non-Dyson approaches have been explored at the $GW$ level, \cite{Bintrim_2021,Tolle_2022,Loos_2026} their behavior within the $G3W2$ framework remains largely unexplored and therefore warrants systematic investigation.
In addition, another natural perspective for this work would be to perform these calculations self-consistently, or include an effective static contribution to the self-energy, which could be used to mimic partial self-consistency effects. 
Although such static corrections have been formulated at the $GW$ level, \cite{Bruneval_2019a,Bruneval_2019b,Bruneval_2021b} their extension to the $G3W2$ approximation remains to be derived.

Beyond these developments, extending the present ADC-based $GW$ and $G3W2$ schemes to the multireference case constitutes another appealing yet challenging direction for future work. 
Multireference extensions within the ADC framework have recently been proposed.\cite{Sokolov_2018,Chatterjee_2019,Chatterjee_2020,Banerjee_2019,Banerjee_2022,Banerjee_2023} 
Although encouraging results have been reported recently, \cite{Wang_2026} the most suitable strategy for incorporating multireference effects into the present formalism remains unclear at this stage.
Together, these perspectives outline several promising avenues for extending the present framework toward a more complete, accurate, and efficient description of electronic correlation within many-body perturbation theory.

Finally, it is worth noting that ADC and coupled-cluster theories \cite{Crawford_2000,Piecuch_2002,Bartlett_2007,Shavitt_2009,Bartlett_2024} share a common diagrammatic language rooted in Goldstone diagrams, which has already enabled the establishment of formal connections between these approaches. \cite{Prasad_1985,Hattig_2005c,Liu_2018,Hodecker_2020a,Hodecker_2020b,Dreuw_2023} 
In this context, the present links between ADC and $GW$ may contribute to a deeper understanding of the relationships between many-body perturbation theory and coupled-cluster methodologies.
Recent studies have further clarified these connections, revealing deep formal links and offering a unified perspective on the respective strengths and limitations of these frameworks.\cite{Lange_2018,Quintero_2022,Tolle_2022,Tolle_2025,Tolle_2025b,Kitsaras_2026,Coveney_2025a,Coveney_2025b,Coveney_2026,Tolle_2026}

\acknowledgements{
This project has received funding from the European Research Council (ERC) under the European Union's Horizon 2020 research and innovation programme (Grant agreement No.~863481).
J.~T.~acknowledges funding from the Fonds der Chemischen Industrie (FCI) via a Liebig fellowship and support by the Cluster of Excellence ``CUI: Advanced Imaging of Matter'' of the Deutsche Forschungsgemeinschaft (DFG) (EXC 2056, funding ID 390715994).
J.~T.~also acknowledges financial support by the Emmy Noether Program of the German Research Foundation (project number  577598009).
This work used the HPC resources from CALMIP (Toulouse) under allocation 2026-18005.}

 
\section*{References}

\bibliography{biblio}

@article{Tolle_2024a,
	author = {T{\"o}lle, Johannes and Kin-Lic Chan, Garnet},
	doi = {10.1063/5.0195934},
	journal = {J. Chem. Phys.},
	month = {04},
	number = {16},
	pages = {164108},
	title = {AB-{{$G_0W_0$}}: A practical {{$G_0W_0$}} method without frequency integration based on an auxiliary boson expansion},
	volume = {160},
	year = {2024}}

@article{Ammar_2024,
	author = {Ammar, Abdallah and Marie, Antoine and {Rodr{\'\i}guez-Mayorga}, Mauricio and Burton, Hugh G. A. and Loos, Pierre-Fran{\c c}ois},
	doi = {10.1063/5.0196561},
	issn = {0021-9606},
	journal = {J. Chem. Phys.},
	month = mar,
	number = {11},
	pages = {114101},
	title = {Can {{$GW$}} Handle Multireference Systems?},
	urldate = {2025-07-03},
	volume = {160},
	year = {2024}}

@article{Coveney_2026,
	author = {Coveney, Christopher J. N. and Tew, David P.},
	doi = {10.1103/9vdr-vhxg},
	issue = {1},
	journal = {Phys. Rev. Res.},
	month = {Mar},
	numpages = {49},
	pages = {013322},
	publisher = {American Physical Society},
	title = {Non-Hermitian {Green's} function theory with {{$N$}}-body interactions: The coupled-cluster similarity transformation},
	url = {https://link.aps.org/doi/10.1103/9vdr-vhxg},
	volume = {8},
	year = {2026}}

@article{Tolle_2025b,
	author = {T{\"o}lle, Johannes and Kitsaras, Marios-Petros and Loos, Pierre-Fran{\c{c}}ois},
	doi = {10.1021/acs.jpclett.5c02219},
	journal = {J. Phys. Chem. Lett.},
	number = {43},
	pages = {11134-11143},
	title = {Fully Analytic Nuclear Gradients for the {Bethe--Salpeter} Equation},
	volume = {16},
	year = {2025}}

@article{Kitsaras_2026,
	author = {Kitsaras, Marios-Petros and T{\"o}lle, Johannes and Loos, Pierre-Fran{\c c}ois},
	doi = {10.1063/5.0309945},
	journal = {J. Chem. Phys.},
	month = {01},
	number = {4},
	pages = {044122},
	title = {Analytic {{$G_0W_0$}} gradients based on a double-similarity transformation equation-of-motion coupled-cluster treatment},
	volume = {164},
	year = {2026}}

@misc{Tolle_2026,
	archiveprefix = {arXiv},
	author = {Johannes T{\"o}lle and Marios-Petros Kitsaras and Andreas Irmler and Andreas Gr{\"u}neis and Pierre-Fran{\c c}ois Loos},
	eprint = {2602.10887},
	primaryclass = {physics.chem-ph},
	title = {Connection between {{$GW$}} and Extended Coupled Cluster},
	url = {https://arxiv.org/abs/2602.10887},
	year = {2026}}

@article{Riva_2023,
	author = {Riva, Gabriele and Romaniello, Pina and Berger, J. Arjan},
	doi = {10.1103/PhysRevLett.131.216401},
	journal = {Phys. Rev. Lett.},
	number = {21},
	pages = {216401},
	title = {Multichannel {{Dyson}} Equation: Coupling Many-Body {{Green}}'s Functions},
	urldate = {2024-04-03},
	volume = {131},
	year = {2023}}

@article{Riva_2024,
	author = {Riva, Gabriele and Romaniello, Pina and Berger, J. Arjan},
	doi = {10.1103/PhysRevB.110.115140},
	journal = {Phys. Rev. B},
	number = {11},
	pages = {115140},
	title = {Derivation and Analysis of the Multichannel {{Dyson}} Equation},
	urldate = {2024-10-15},
	volume = {110},
	year = {2024}}

@misc{Romaniello_2026,
	archiveprefix = {arXiv},
	author = {Pina Romaniello and J. Arjan Berger},
	eprint = {2603.27329},
	title = {Direct and inverse photoemission spectra from the screened multichannel Dyson equation},
	url = {https://arxiv.org/abs/2603.27329},
	year = {2026}}

@misc{Loos_2026,
	archiveprefix = {arXiv},
	author = {Pierre-Fran{\c c}ois Loos and Johannes T{\"o}lle},
	eprint = {2604.08350},
	primaryclass = {physics.chem-ph},
	title = {From Full Dynamic to Pure Static: A Family of {{$GW$}}-Based Approximations},
	url = {https://arxiv.org/abs/2604.08350},
	year = {2026}}

@article{Hodecker_2020b,
	author = {Hodecker, Manuel and Thielen, Sebastian M. and Liu, Junzi and Rehn, Dirk R. and Dreuw, Andreas},
	doi = {10.1021/acs.jctc.0c00335},
	journal = {J. Chem. Theory Comput.},
	number = {6},
	pages = {3654-3663},
	title = {Third-Order Unitary Coupled Cluster {(UCC3)} for Excited Electronic States: Efficient Implementation and Benchmarking},
	volume = {16},
	year = {2020}}

@article{Liu_2018,
	author = {Liu, Junzi and Asthana, Ayush and Cheng, Lan and Mukherjee, Debashis},
	doi = {10.1063/1.5030344},
	journal = {J. Chem. Phys.},
	month = {06},
	number = {24},
	pages = {244110},
	title = {Unitary coupled-cluster based self-consistent polarization propagator theory: A third-order formulation and pilot applications},
	volume = {148},
	year = {2018}}

@article{Prasad_1985,
	author = {Prasad, M. Durga and Pal, Sourav and Mukherjee, Debashis},
	doi = {10.1103/PhysRevA.31.1287},
	issue = {3},
	journal = {Phys. Rev. A},
	month = {Mar},
	numpages = {0},
	pages = {1287--1298},
	publisher = {American Physical Society},
	title = {Some aspects of self-consistent propagator theories},
	url = {https://link.aps.org/doi/10.1103/PhysRevA.31.1287},
	volume = {31},
	year = {1985}}

@article{Hodecker_2020a,
	author = {Hodecker, Manuel and Rehn, Dirk R. and Dreuw, Andreas},
	doi = {10.1063/1.5142354},
	journal = {J. Chem. Phys.},
	month = {03},
	number = {9},
	pages = {094106},
	title = {Hermitian second-order methods for excited electronic states: Unitary coupled cluster in comparison with algebraic--diagrammatic construction schemes},
	volume = {152},
	year = {2020}}

@misc{Wang_2026,
	archiveprefix = {arXiv},
	author = {Yuqi Wang and Wei-Hai Fang and Zhendong Li},
	eprint = {2604.16013},
	primaryclass = {physics.chem-ph},
	title = {Multi-reference {{$GW$}} approximation for strongly correlated molecules},
	url = {https://arxiv.org/abs/2604.16013},
	year = {2026}}

@article{Bruneval_2019a,
	author = {Bruneval, Fabien},
	doi = {10.1103/PhysRevB.99.041118},
	journal = {Phys. Rev. B},
	number = {4},
	pages = {041118},
	title = {Improved density matrices for accurate molecular ionization potentials},
	volume = {99},
	year = {2019}}

@article{Bruneval_2019b,
	author = {Bruneval, Fabien},
	doi = {10.1021/acs.jctc.9b00333},
	journal = {J. Chem. Theory Comput.},
	number = {7},
	pages = {4069--4078},
	title = {Assessment of the linearized {{$GW$}} density matrix for molecules},
	volume = {15},
	year = {2019}}

@article{Bruneval_2021b,
	author = {Bruneval, Fabien and Rodriguez-Mayorga, Mauricio and Rinke, Patrick and Dvorak, Marc},
	doi = {10.1021/acs.jctc.0c01264},
	journal = {J. Chem. Theory Comput.},
	number = {4},
	pages = {2126--2136},
	title = {Improved one-shot total energies from the linearized {{$GW$}} density matrix},
	volume = {17},
	year = {2021}}

@article{Marie_2025b,
	author = {Marie, Antoine and Loos, Pierre-Fran{\c c}ois},
	doi = {10.1063/5.0301331},
	journal = {J. Chem. Phys.},
	month = {11},
	number = {19},
	pages = {194115},
	title = {Parquet theory for molecular systems: Formalism and static kernel parquet approximation},
	volume = {163},
	year = {2025}}

@article{Sun_2020,
	author = {Sun, Qiming and Zhang, Xing and Banerjee, Samragni and Bao, Peng and Barbry, Marc and Blunt, Nick S. and Bogdanov, Nikolay A. and Booth, George H. and Chen, Jia and Cui, Zhi-Hao and Eriksen, Janus J. and Gao, Yang and Guo, Sheng and Hermann, Jan and Hermes, Matthew R. and Koh, Kevin and Koval, Peter and Lehtola, Susi and Li, Zhendong and Liu, Junzi and Mardirossian, Narbe and McClain, James D. and Motta, Mario and Mussard, Bastien and Pham, Hung Q. and Pulkin, Artem and Purwanto, Wirawan and Robinson, Paul J. and Ronca, Enrico and Sayfutyarova, Elvira R. and Scheurer, Maximilian and Schurkus, Henry F. and Smith, James E. T. and Sun, Chong and Sun, Shi-Ning and Upadhyay, Shiv and Wagner, Lucas K. and Wang, Xiao and White, Alec and Whitfield, James Daniel and Williamson, Mark J. and Wouters, Sebastian and Yang, Jun and Yu, Jason M. and Zhu, Tianyu and Berkelbach, Timothy C. and Sharma, Sandeep and Sokolov, Alexander Yu. and Chan, Garnet Kin-Lic},
	doi = {10.1063/5.0006074},
	journal = {J. Chem. Phys.},
	month = {07},
	number = {2},
	pages = {024109},
	title = {Recent developments in the PySCF program package},
	volume = {153},
	year = {2020}}

@article{Sun_2018,
	author = {Sun, Qiming and Berkelbach, Timothy C. and Blunt, Nick S. and Booth, George H. and Guo, Sheng and Li, Zhendong and Liu, Junzi and McClain, James D. and Sayfutyarova, Elvira R. and Sharma, Sandeep and Wouters, Sebastian and Chan, Garnet Kin-Lic},
	doi = {https://doi.org/10.1002/wcms.1340},
	journal = {WIREs Comput. Mol. Sci.},
	number = {1},
	pages = {e1340},
	title = {PySCF: the Python-based simulations of chemistry framework},
	volume = {8},
	year = {2018}}

@article{Cederbaum_1975,
	author = {Cederbaum, L. S.},
	doi = {10.1063/1.430783},
	journal = {J. Chem. Phys.},
	month = {03},
	number = {6},
	pages = {2160-2170},
	title = {Non-single-particle excitations in finite {Fermi} systems},
	volume = {62},
	year = {1975}}

@article{Winter_1972,
	author = {J. Winter},
	doi = {10.1016/0375-9474(72)91000-7},
	journal = {Nuc. Phys. A},
	number = {3},
	pages = {535-551},
	title = {Study of core excitations in one-particle and one-holf nuclei by means of the six-point {Green} function},
	volume = {194},
	year = {1972}}

@article{Uimonen_2015,
	author = {Uimonen, A.-M. and Stefanucci, G. and Pavlyukh, Y. and van Leeuwen, R.},
	doi = {10.1103/PhysRevB.91.115104},
	issue = {11},
	journal = {Phys. Rev. B},
	month = {Mar},
	numpages = {18},
	pages = {115104},
	publisher = {American Physical Society},
	title = {Diagrammatic expansion for positive density-response spectra: Application to the electron gas},
	url = {https://link.aps.org/doi/10.1103/PhysRevB.91.115104},
	volume = {91},
	year = {2015}}

@article{Stefanucci_2014,
	author = {Stefanucci, G. and Pavlyukh, Y. and Uimonen, A.-M. and van Leeuwen, R.},
	doi = {10.1103/physrevb.90.115134},
	journal = {Phys. Rev. B},
	pages = {115134},
	title = {Diagrammatic expansion for positive spectral functions beyond {$GW$}: Application to vertex corrections in the electron gas},
	volume = {90},
	year = {2014}}

@article{Patterson_2024,
	author = {Patterson, Charles H.},
	doi = {10.1021/acs.jctc.4c00795},
	journal = {J. Chem. Theory Comput.},
	number = {17},
	pages = {7479-7493},
	title = {Molecular Ionization Energies from {{$GW$}} and {Hartree--Fock} Theory: Polarizability, Screening, and Self-Energy Vertex Corrections},
	volume = {20},
	year = {2024}}

@article{Vacondio_2024,
	author = {Vacondio, Simone and Varsano, Daniele and Ruini, Alice and Ferretti, Andrea},
	doi = {10.1021/acs.jctc.4c00100},
	journal = {J. Chem. Theory Comput.},
	number = {11},
	pages = {4718-4737},
	title = {Going Beyond the {{$GW$}} Approximation Using the Time-Dependent {{Hartree--Fock}} Vertex},
	volume = {20},
	year = {2024}}

@article{Kuwahara_2016,
	author = {Kuwahara, Riichi and Noguchi, Yoshifumi and Ohno, Kaoru},
	doi = {10.1103/PhysRevB.94.121116},
	issue = {12},
	journal = {Phys. Rev. B},
	month = {Sep},
	numpages = {5},
	pages = {121116},
	publisher = {American Physical Society},
	title = {{{$GW\Gamma$+}} {Bethe-Salpeter} equation approach for photoabsorption spectra: Importance of self-consistent {{$GW\Gamma$}} calculations in small atomic systems},
	url = {https://link.aps.org/doi/10.1103/PhysRevB.94.121116},
	volume = {94},
	year = {2016}}

@article{Rohlfing_2023,
	author = {Rohlfing, Michael},
	doi = {10.1103/PhysRevB.108.195207},
	issue = {19},
	journal = {Phys. Rev. B},
	month = {Nov},
	numpages = {33},
	pages = {195207},
	publisher = {American Physical Society},
	title = {Approximate spatiotemporal structure of the vertex function $\mathrm{\ensuremath{\Gamma}}(1, 2; 3)$ in many-body perturbation theory},
	url = {https://link.aps.org/doi/10.1103/PhysRevB.108.195207},
	volume = {108},
	year = {2023}}

@article{Tal_2021,
	author = {Tal, Alexey and Chen, Wei and Pasquarello, Alfredo},
	doi = {10.1103/PhysRevB.103.L161104},
	issue = {16},
	journal = {Phys. Rev. B},
	month = {Apr},
	numpages = {5},
	pages = {L161104},
	publisher = {American Physical Society},
	title = {Vertex function compliant with the Ward identity for quasiparticle self-consistent calculations beyond {{$GW$}}},
	url = {https://link.aps.org/doi/10.1103/PhysRevB.103.L161104},
	volume = {103},
	year = {2021}}

@article{Ma_2019,
	author = {Ma, He and Govoni, Marco and Gygi, Francois and Galli, Giulia},
	doi = {10.1021/acs.jctc.8b00864},
	journal = {J. Chem. Theory Comput.},
	number = {1},
	pages = {154-164},
	title = {A Finite-Field Approach for {{$GW$}} Calculations beyond the Random Phase Approximation},
	volume = {15},
	year = {2019}}

@article{Pavlyukh_2020,
	author = {Pavlyukh, Y. and Stefanucci, G. and van Leeuwen, R.},
	doi = {10.1103/PhysRevB.102.045121},
	issue = {4},
	journal = {Phys. Rev. B},
	month = {Jul},
	numpages = {22},
	pages = {045121},
	publisher = {American Physical Society},
	title = {Dynamically screened vertex correction to {{$GW$}}},
	url = {https://link.aps.org/doi/10.1103/PhysRevB.102.045121},
	volume = {102},
	year = {2020}}

@article{Cunningham_2023,
	author = {Cunningham, Brian and Gr\"uning, Myrta and Pashov, Dimitar and van Schilfgaarde, Mark},
	doi = {10.1103/PhysRevB.108.165104},
	issue = {16},
	journal = {Phys. Rev. B},
	month = {Oct},
	numpages = {39},
	pages = {165104},
	publisher = {American Physical Society},
	title = {{{QS$GW$}}: Quasiparticle self-consistent {{$GW$ with ladder diagrams in $W$}}},
	url = {https://link.aps.org/doi/10.1103/PhysRevB.108.165104},
	volume = {108},
	year = {2023}}

@article{Schmidt_2017,
	author = {Schmidt, Per S. and Patrick, Christopher E. and Thygesen, Kristian S.},
	doi = {10.1103/PhysRevB.96.205206},
	issue = {20},
	journal = {Phys. Rev. B},
	month = {Nov},
	numpages = {9},
	pages = {205206},
	publisher = {American Physical Society},
	title = {Simple vertex correction improves {{$GW$}} band energies of bulk and two-dimensional crystals},
	url = {https://link.aps.org/doi/10.1103/PhysRevB.96.205206},
	volume = {96},
	year = {2017}}

@article{Kutepov_2021,
	author = {Kutepov, A L},
	doi = {10.1088/1361-648X/ac23fa},
	journal = {J. Phys. Cond. Mat.},
	month = {sep},
	number = {48},
	pages = {485601},
	title = {Spatial non-locality of electronic correlations beyond {{$GW$}} approximation},
	volume = {33},
	year = {2021}}

@article{Bruneval_2005,
	author = {Bruneval, Fabien and Sottile, Francesco and Olevano, Valerio and Del Sole, Rodolfo and Reining, Lucia},
	doi = {10.1103/PhysRevLett.94.186402},
	issue = {18},
	journal = {Phys. Rev. Lett.},
	month = {May},
	numpages = {4},
	pages = {186402},
	publisher = {American Physical Society},
	title = {Many-Body Perturbation Theory Using the Density-Functional Concept: Beyond the {{$GW$}} Approximation},
	url = {https://link.aps.org/doi/10.1103/PhysRevLett.94.186402},
	volume = {94},
	year = {2005}}

@article{Northrup_1987,
	author = {Northrup, John E. and Hybertsen, Mark S. and Louie, Steven G.},
	doi = {10.1103/PhysRevLett.59.819},
	issue = {7},
	journal = {Phys. Rev. Lett.},
	month = {Aug},
	numpages = {0},
	pages = {819--822},
	publisher = {American Physical Society},
	title = {Theory of quasiparticle energies in alkali metals},
	url = {https://link.aps.org/doi/10.1103/PhysRevLett.59.819},
	volume = {59},
	year = {1987}}

@article{Tomczak_2017,
	author = {Tomczak, J. M. and Liu, P. and Toschi, A. and Kresse, G. and Held, K.},
	doi = {10.1140/epjst/e2017-70053-1},
	issn = {1951-6401},
	journal = {Eur. Phys. J. Spec. Top.},
	month = jul,
	number = {11},
	pages = {2565--2590},
	publisher = {Springer Science and Business Media LLC},
	title = {Merging {{$GW$ with DMFT}} and non-local correlations beyond},
	url = {http://dx.doi.org/10.1140/epjst/e2017-70053-1},
	volume = {226},
	year = {2017}}

@article{DiSabatino_2022,
	author = {Di Sabatino, S. and Koskelo, J. and Berger, J. A. and Romaniello, P.},
	doi = {10.1103/PhysRevB.105.235123},
	issue = {23},
	journal = {Phys. Rev. B},
	month = {Jun},
	numpages = {11},
	pages = {235123},
	publisher = {American Physical Society},
	title = {Introducing screening in one-body density matrix functionals: Impact on charged excitations of model systems via the extended {Koopmans'} theorem},
	url = {https://link.aps.org/doi/10.1103/PhysRevB.105.235123},
	volume = {105},
	year = {2022}}

@article{DiSabatino_2023,
	author = {Di Sabatino, S. and Koskelo, J. and Berger, J. A. and Romaniello, P.},
	doi = {10.1103/PhysRevB.107.035111},
	issue = {3},
	journal = {Phys. Rev. B},
	month = {Jan},
	numpages = {5},
	pages = {035111},
	publisher = {American Physical Society},
	title = {Screened extended {Koopmans'} theorem: Photoemission at weak and strong correlation},
	url = {https://link-aps-org-s.docadis.univ-tlse3.fr/doi/10.1103/PhysRevB.107.035111},
	volume = {107},
	year = {2023}}

@article{Mertins_1996b,
	author = {Mertins, F. and Schirmer, J. and Tarantelli, A.},
	doi = {10.1103/PhysRevA.53.2153},
	issue = {4},
	journal = {Phys. Rev. A},
	month = {Apr},
	numpages = {0},
	pages = {2153--2168},
	publisher = {American Physical Society},
	title = {Algebraic propagator approaches and intermediate-state representations. {II. The} equation-of-motion methods for {N, N\ifmmode\pm\else\textpm\fi{}1, and N\ifmmode\pm\else\textpm\fi{}2} electrons},
	url = {https://link.aps.org/doi/10.1103/PhysRevA.53.2153},
	volume = {53},
	year = {1996}}

@article{Mertins_1996a,
	author = {Mertins, F. and Schirmer, J.},
	doi = {10.1103/PhysRevA.53.2140},
	issue = {4},
	journal = {Phys. Rev. A},
	month = {Apr},
	numpages = {0},
	pages = {2140--2152},
	publisher = {American Physical Society},
	title = {Algebraic propagator approaches and intermediate-state representations. {I. The} biorthogonal and unitary coupled-cluster methods},
	url = {https://link.aps.org/doi/10.1103/PhysRevA.53.2140},
	volume = {53},
	year = {1996}}

@article{Schirmer_1998,
	author = {Schirmer, J. and Trofimov, A. B. and Stelter, G.},
	doi = {10.1063/1.477085},
	journal = {J. Chem. Phys.},
	month = {09},
	number = {12},
	pages = {4734-4744},
	title = {A non-{Dyson} third-order approximation scheme for the electron propagator},
	volume = {109},
	year = {1998}}

@article{vonNiessen_1984,
	author = {von Niessen, W. and Schirmer, J. and Cederbaum, L.S.},
	doi = {10.1016/0167-7977(84)90002-9},
	issn = {0167-7977},
	journal = {Comput. Phys. Rep.},
	month = apr,
	number = {2},
	pages = {57--125},
	publisher = {Elsevier BV},
	title = {Computational methods for the one-particle Green's function},
	url = {http://dx.doi.org/10.1016/0167-7977(84)90002-9},
	volume = {1},
	year = {1984}}

@article{Dreuw_2023,
	author = {Dreuw, Andreas and Papapostolou, Antonia and Dempwolff, Adrian L.},
	doi = {10.1021/acs.jpca.3c02761},
	journal = {J. Phys. Chem. A},
	number = {32},
	pages = {6635-6646},
	title = {Algebraic Diagrammatic Construction Schemes Employing the Intermediate State Formalism: Theory, Capabilities, and Interpretation},
	volume = {127},
	year = {2023}}

@article{Banerjee_2022,
	author = {Banerjee, Samragni and Sokolov, Alexander Yu.},
	doi = {10.1021/acs.jctc.2c00565},
	journal = {J. Chem. Theory Comput.},
	number = {9},
	pages = {5337-5348},
	title = {Non-{Dyson} Algebraic Diagrammatic Construction Theory for Charged Excitations in Solids},
	volume = {18},
	year = {2022}}

@article{Loos_2024,
	author = {Loos, Pierre-Fran{\c c}ois and Marie, Antoine and Ammar, Abdallah},
	doi = {10.1039/D4FD00037D},
	issue = {0},
	journal = {Faraday Discuss.},
	pages = {240-260},
	publisher = {The Royal Society of Chemistry},
	title = {Cumulant {Green's} function methods for molecules},
	url = {http://dx.doi.org/10.1039/D4FD00037D},
	volume = {254},
	year = {2024}}

@article{Bruneval_2025,
	author = {Bruneval, Fabien and F{\"o}rster, Arno and Pavlyukh, Yaroslav},
	doi = {10.1021/acs.jctc.5c01180},
	journal = {J. Chem. Theory Comput.},
	number = {20},
	pages = {10223-10240},
	title = {{{$GW$+2SOSEX}} Self-Energy Made Positive Semidefinite},
	volume = {21},
	year = {2025}}

@article{Forster_2025,
	author = {F{\"o}rster, Arno},
	doi = {10.1021/acs.jctc.4c01639},
	journal = {J. Chem. Theory Comput.},
	number = {4},
	pages = {1709-1721},
	title = {Beyond Quasi-Particle Self-Consistent {{$GW$}} for Molecules with Vertex Corrections},
	volume = {21},
	year = {2025}}

@article{Forster_2024,
	author = {F{\"o}rster, Arno and Bruneval, Fabien},
	doi = {10.1021/acs.jpclett.4c03126},
	journal = {J. Phys. Chem. Lett.},
	number = {51},
	pages = {12526-12534},
	title = {Why Does the {{$GW$}} Approximation Give Accurate Quasiparticle Energies? {{The}} Cancellation of Vertex Corrections Quantified},
	volume = {15},
	year = {2024}}

@article{Bruneval_2024,
	author = {Bruneval, Fabien and F{\"o}rster, Arno},
	doi = {10.1021/acs.jctc.4c00090},
	issn = {1549-9618},
	journal = {J. Chem. Theory Comput.},
	number = {8},
	pages = {3218--3230},
	title = {Fully Dynamic {{$G3W2$}} Self-Energy for Finite Systems: Formulas and Benchmark},
	urldate = {2024-05-16},
	volume = {20},
	year = {2024}}

@article{Wen_2024,
	author = {Wen, Ming and Abraham, Vibin and Harsha, Gaurav and Shee, Avijit and Whaley, K. Birgitta and Zgid, Dominika},
	copyright = {https://doi.org/10.15223/policy-029},
	doi = {10.1021/acs.jctc.3c01279},
	issn = {1549-9618, 1549-9626},
	journal = {J. Chem. Theo. Comput.},
	number = {8},
	pages = {3109--3120},
	publisher = {American Chemical Society (ACS)},
	title = {Comparing Self-Consistent {{$GW$}} and Vertex-Corrected {{$G_0W_0$}} ({{$G_0W_0\Gamma$}}) Accuracy for Molecular Ionization Potentials},
	urldate = {2025-07-25},
	volume = {20},
	year = {2024}}

@article{Weng_2023,
	author = {Weng, Guorong and Mallarapu, Rushil and Vl{\v c}ek, Vojt{\v e}ch},
	doi = {10.1063/5.0139117},
	issn = {0021-9606},
	journal = {. Chem. Phys.},
	number = {14},
	pages = {144105},
	title = {Embedding Vertex Corrections in {{GW}} Self-Energy: {{Theory}}, Implementation, and Outlook},
	urldate = {2024-03-26},
	volume = {158},
	year = {2023}}

@article{Forster_2022a,
	author = {F{\"o}rster, Arno and Visscher, Lucas},
	doi = {10.1103/PhysRevB.105.125121},
	journal = {Phys. Rev. B},
	number = {12},
	pages = {125121},
	title = {Exploring the Statically Screened {{$G_3W_2$}} Correction to the {{$GW$}} Self-Energy: {{Charged}} Excitations and Total Energies of Finite Systems},
	urldate = {2024-05-27},
	volume = {105},
	year = {2022}}

@article{Wang_2021a,
	author = {Wang, Yanyong and Rinke, Patrick and Ren, Xinguo},
	doi = {10.1021/acs.jctc.1c00488},
	issn = {1549-9618},
	journal = {J. Chem. Theory Comput.},
	number = {8},
	pages = {5140--5154},
	title = {Assessing the {{$G_0W_0\Gamma_0$}}(1) {{Approach}}: {{Beyond $G_0W_0$}} with {{Hedin}}'s {{Full Second-Order Self-Energy Contribution}}},
	urldate = {2024-05-27},
	volume = {17},
	year = {2021}}

@article{Bartlett_2024,
	author = {Bartlett, Rodney J.},
	doi = {10.1039/D3CP03853J},
	issue = {10},
	journal = {Phys. Chem. Chem. Phys.},
	pages = {8013-8037},
	publisher = {The Royal Society of Chemistry},
	title = {Perspective on Coupled-cluster Theory. {The} evolution toward simplicity in quantum chemistry},
	url = {http://dx.doi.org/10.1039/D3CP03853J},
	volume = {26},
	year = {2024}}

@article{Scott_2023,
	author = {Scott, Charles J. C. and Backhouse, Oliver J. and Booth, George H.},
	doi = {10.1063/5.0143291},
	journal = {J. Chem. Phys.},
	month = {03},
	number = {12},
	pages = {124102},
	title = {A ``moment-conserving'' reformulation of {{$GW$}} theory},
	volume = {158},
	year = {2023}}

@article{Coveney_2025a,
	author = {Coveney, Christopher J. N. and Tew, David P.},
	doi = {10.1103/p41w-bl6p},
	issue = {4},
	journal = {Phys. Rev. B},
	month = {Jul},
	numpages = {17},
	pages = {045104},
	publisher = {American Physical Society},
	title = {Diagrammatic theory of the irreducible coupled-cluster self-energy},
	url = {https://link.aps.org/doi/10.1103/p41w-bl6p},
	volume = {112},
	year = {2025}}

@article{Coveney_2025b,
	author = {Coveney, Christopher J. N.},
	doi = {10.1021/acs.jpca.5c03750},
	journal = {J. Phys. Chem. A},
	number = {37},
	pages = {8689-8698},
	title = {Uncovering Relationships between the Electronic Self-Energy and Coupled-Cluster Doubles Theory},
	volume = {129},
	year = {2025}}

@article{Shirley_1993,
	author = {Shirley, Eric L. and Martin, Richard M.},
	doi = {10.1103/PhysRevB.47.15404},
	issue = {23},
	journal = {Phys. Rev. B},
	month = {Jun},
	numpages = {0},
	pages = {15404--15412},
	publisher = {American Physical Society},
	title = {{{$GW$}} quasiparticle calculations in atoms},
	url = {https://link.aps.org/doi/10.1103/PhysRevB.47.15404},
	volume = {47},
	year = {1993}}

@article{Tolle_2024b,
	author = {T{\"o}lle, Johannes and Niemeyer, Niklas and Neugebauer, Johannes},
	doi = {10.1021/acs.jctc.3c01264},
	journal = {J. Chem. Theory Comput.},
	number = {5},
	pages = {2022-2032},
	title = {Accelerating Analytic-Continuation {{$GW$}} Calculations with a {Laplace} Transform and Natural Auxiliary Functions},
	volume = {20},
	year = {2024}}

@article{Yu_2022,
	author = {Yu, Victor Wen-Zhe and Govoni, Marco},
	doi = {10.1021/acs.jctc.2c00241},
	journal = {J. Chem. Theory Comput.},
	number = {8},
	pages = {4690-4707},
	title = {{GPU Acceleration of Large-Scale Full-Frequency $GW$ Calculations}},
	volume = {18},
	year = {2022}}

@article{Wilhelm_2021,
	author = {Wilhelm, Jan and Seewald, Patrick and Golze, Dorothea},
	doi = {10.1021/acs.jctc.0c01282},
	journal = {J. Chem. Theory Comput.},
	number = {3},
	pages = {1662-1677},
	title = {{Low-Scaling $GW$ with Benchmark Accuracy and Application to Phosphorene Nanosheets}},
	volume = {17},
	year = {2021}}

@article{Kaltak_2020,
	author = {Kaltak, Merzuk and Kresse, Georg},
	doi = {10.1103/PhysRevB.101.205145},
	issue = {20},
	journal = {Phys. Rev. B},
	month = {May},
	numpages = {17},
	pages = {205145},
	publisher = {American Physical Society},
	title = {Minimax isometry method: A compressive sensing approach for {Matsubara} summation in many-body perturbation theory},
	url = {https://link.aps.org/doi/10.1103/PhysRevB.101.205145},
	volume = {101},
	year = {2020}}

@article{Mukatayev_2023,
	author = {Mukatayev, Iskander and Moevus, Florient and Skl{\'e}nard, Beno{\^\i}t and Olevano, Valerio and Li, Jing},
	doi = {10.1021/acs.jpca.3c00173},
	issn = {1089-5639},
	journal = {J. Phys. Chem. A},
	number = {7},
	pages = {1642--1648},
	title = {{{XPS Core-Level Chemical Shift}} by {{Ab Initio Many-Body Theory}}},
	urldate = {2025-01-08},
	volume = {127},
	year = {2023}}

@article{Mejia-Rodriguez_2021,
	author = {{Mejia-Rodriguez}, Daniel and Kunitsa, Alexander and Apr{\`a}, Edoardo and Govind, Niranjan},
	doi = {10.1021/acs.jctc.1c00738},
	issn = {1549-9618},
	journal = {J. Chem. Theory Comput.},
	number = {12},
	pages = {7504--7517},
	title = {Scalable {{Molecular $GW$ Calculations}}: {{Valence}} and {{Core Spectra}}},
	urldate = {2025-01-08},
	volume = {17},
	year = {2021}}

@incollection{Marie_2024a,
	author = {Marie, Antoine and Ammar, Abdallah and Loos, Pierre-Fran{\c c}ois},
	booktitle = {Advances in {{Quantum Chemistry}}},
	doi = {10.1016/bs.aiq.2024.04.001},
	pages = {157--184},
	series = {Novel {{Treatments}} of {{Strong Correlations}}},
	title = {The {{$GW$}} Approximation: {{A}} Quantum Chemistry Perspective},
	urldate = {2024-10-08},
	volume = {90},
	year = {2024}}

@article{Marie_2024b,
	author = {Marie, Antoine and Loos, Pierre-Fran{\c c}ois},
	doi = {10.1021/acs.jctc.4c00216},
	issn = {1549-9618},
	journal = {J. Chem. Theory Comput.},
	number = {11},
	pages = {4751--4777},
	title = {Reference {{Energies}} for {{Valence Ionizations}} and {{Satellite Transitions}}},
	urldate = {2024-10-15},
	volume = {20},
	year = {2024}}

@article{Tolle_2025,
	author = {T{\"o}lle, Johannes},
	doi = {10.1021/acs.jpclett.5c00046},
	journal = {J. Phys. Chem. Lett.},
	number = {15},
	pages = {3672-3678},
	title = {Fully Analytic {{$G_0W_0$}} Nuclear Gradients},
	volume = {16},
	year = {2025}}

@article{Quintero_2022,
	author = {Quintero-Monsebaiz, Ra{\'u}l and Monino, Enzo and Marie, Antoine and Loos, Pierre-Fran{\c c}ois},
	doi = {10.1063/5.0130837},
	journal = {J. Chem. Phys.},
	month = {12},
	number = {23},
	pages = {231102},
	title = {{Connections between many-body perturbation and coupled-cluster theories}},
	volume = {157},
	year = {2022}}

@article{Monino_2023,
	author = {Enzo Monino and Pierre-Fran{\c{c}}ois Loos},
	doi = {10.1063/5.0159853},
	journal = {J. Chem. Phys.},
	month = jul,
	number = {3},
	pages = {034105},
	publisher = {{AIP} Publishing},
	title = {Connections and performances of Green's function methods for charged and neutral excitations},
	url = {https://doi.org/10.1063/5.0159853},
	volume = {159},
	year = {2023}}

@article{Hedin_1965,
	author = {Hedin, Lars},
	doi = {10.1103/PhysRev.139.A796},
	journal = {Phys. Rev.},
	number = {3A},
	pages = {A796},
	title = {New Method for Calculating the One-Particle {{Green}}'s Function with Application to the Electron-Gas Problem},
	volume = {139},
	year = {1965}}

@article{Orlando_2023b,
	author = {Orlando, Roberto and Romaniello, Pina and Loos, Pierre-Fran{\c c}ois},
	doi = {10.1063/5.0176898},
	journal = {J. Chem. Phys.},
	month = {11},
	number = {18},
	pages = {184113},
	title = {{The three channels of many-body perturbation theory: {{$GW$}}, particle--particle, and electron--hole {{$T$}}-matrix self-energies}},
	volume = {159},
	year = {2023}}

@article{Sokolov_2018,
	author = {Sokolov, Alexander Yu.},
	doi = {10.1063/1.5055380},
	issn = {0021-9606},
	journal = {J. Chem. Phys.},
	number = {20},
	pages = {204113},
	title = {Multi-Reference Algebraic Diagrammatic Construction Theory for Excited States: {{General}} Formulation and First-Order Implementation},
	urldate = {2024-01-24},
	volume = {149},
	year = {2018}}

@article{Banerjee_2019,
	author = {Banerjee, Samragni and Sokolov, Alexander Yu.},
	doi = {10.1063/1.5131771},
	issn = {0021-9606},
	journal = {J. Chem. Phys.},
	number = {22},
	pages = {224112},
	title = {Third-Order Algebraic Diagrammatic Construction Theory for Electron Attachment and Ionization Energies: {{Conventional}} and {{Green}}'s Function Implementation},
	urldate = {2023-12-19},
	volume = {151},
	year = {2019}}

@article{Banerjee_2023,
	author = {Banerjee, Samragni and Sokolov, Alexander Yu.},
	doi = {10.1021/acs.jctc.3c00251},
	issn = {1549-9618},
	journal = {J. Chem. Theory Comput.},
	number = {11},
	pages = {3037--3053},
	title = {Algebraic {{Diagrammatic Construction Theory}} for {{Simulating Charged Excited States}} and {{Photoelectron Spectra}}},
	urldate = {2023-12-19},
	volume = {19},
	year = {2023}}

@article{Forster_2022,
	author = {F{\"o}rster, Arno and Visscher, Lucas},
	doi = {10.1021/acs.jctc.2c00531},
	issn = {1549-9618},
	journal = {J. Chem. Theory Comput.},
	number = {11},
	pages = {6779--6793},
	title = {Quasiparticle {{Self-Consistent $GW$-Bethe}}{\textendash}{{Salpeter Equation Calculations}} for {{Large Chromophoric Systems}}},
	urldate = {2023-12-19},
	volume = {18},
	year = {2022}}

@article{Chatterjee_2019,
	author = {Chatterjee, Koushik and Sokolov, Alexander Yu.},
	doi = {10.1021/acs.jctc.9b00528},
	issn = {1549-9618},
	journal = {J. Chem. Theory Comput.},
	number = {11},
	pages = {5908--5924},
	title = {Second-{{Order Multireference Algebraic Diagrammatic Construction Theory}} for {{Photoelectron Spectra}} of {{Strongly Correlated Systems}}},
	urldate = {2023-12-19},
	volume = {15},
	year = {2019}}

@article{Chatterjee_2020,
	author = {Chatterjee, Koushik and Sokolov, Alexander Yu.},
	doi = {10.1021/acs.jctc.0c00778},
	issn = {1549-9618},
	journal = {J. Chem. Theory Comput.},
	number = {10},
	pages = {6343--6357},
	title = {Extended {{Second-Order Multireference Algebraic Diagrammatic Construction Theory}} for {{Charged Excitations}}},
	urldate = {2023-12-19},
	volume = {16},
	year = {2020}}

@article{Marie_2023,
	author = {Marie, Antoine and Loos, Pierre-Fran{\c c}ois},
	doi = {10.1021/acs.jctc.3c00281},
	issn = {1549-9618},
	journal = {J. Chem. Theory Comput.},
	number = {13},
	pages = {3943--3957},
	title = {A {{Similarity Renormalization Group Approach}} to {{Green}}'s {{Function Methods}}},
	urldate = {2023-12-04},
	volume = {19},
	year = {2023}}

@article{Govoni_2015,
	author = {Marco Govoni and Giulia Galli},
	doi = {10.1021/ct500958p},
	journal = {J. Chem. Theory Comput.},
	month = may,
	number = {6},
	pages = {2680--2696},
	publisher = {American Chemical Society ({ACS})},
	title = {Large Scale {{$GW$}} Calculations},
	url = {https://doi.org/10.1021/ct500958p},
	volume = {11},
	year = {2015}}

@article{DelBen_2019,
	author = {Mauro Del Ben and Felipe H. da Jornada and Andrew Canning and Nathan Wichmann and Karthik Raman and Ruchira Sasanka and Chao Yang and Steven G. Louie and Jack Deslippe},
	doi = {10.1016/j.cpc.2018.09.003},
	journal = {Comp. Phys. Comm.},
	month = feb,
	pages = {187--195},
	publisher = {Elsevier {BV}},
	title = {Large-scale {{$GW$}} calculations on pre-exascale {HPC} systems},
	url = {https://doi.org/10.1016/j.cpc.2018.09.003},
	volume = {235},
	year = {2019}}

@article{Panades-Barrueta_2023,
	author = {Panad{\'e}s-Barrueta, Ram{\'o}n L. and Golze, Dorothea},
	doi = {10.1021/acs.jctc.3c00555},
	journal = {J. Chem. Theory Comput.},
	number = {16},
	pages = {5450-5464},
	title = {Accelerating Core-Level {{$GW$}} Calculations by Combining the Contour Deformation Approach with the Analytic Continuation of {{$W$}}},
	volume = {19},
	year = {2023}}

@article{Shirley_1996,
	author = {Shirley, Eric L.},
	doi = {10.1103/PhysRevB.54.7758},
	issue = {11},
	journal = {Phys. Rev. B},
	month = {Sep},
	numpages = {0},
	pages = {7758--7764},
	publisher = {American Physical Society},
	title = {Self-consistent {{$GW$}} and higher-order calculations of electron states in metals},
	url = {https://link.aps.org/doi/10.1103/PhysRevB.54.7758},
	volume = {54},
	year = {1996}}

@article{Maggio_2017b,
	author = {Maggio, Emanuele and Kresse, Georg},
	doi = {10.1021/acs.jctc.7b00586},
	journal = {J. Chem. Theory Comput.},
	number = {10},
	pages = {4765-4778},
	title = {{{$GW$}} Vertex Corrected Calculations for Molecular Systems},
	volume = {13},
	year = {2017}}

@article{Gruneis_2014,
	author = {Gr\"uneis, Andreas and Kresse, Georg and Hinuma, Yoyo and Oba, Fumiyasu},
	doi = {10.1103/PhysRevLett.112.096401},
	issue = {9},
	journal = {Phys. Rev. Lett.},
	month = {Mar},
	numpages = {5},
	pages = {096401},
	publisher = {American Physical Society},
	title = {Ionization Potentials of Solids: The Importance of Vertex Corrections},
	url = {https://link.aps.org/doi/10.1103/PhysRevLett.112.096401},
	volume = {112},
	year = {2014}}

@article{Shishkin_2007b,
	author = {Shishkin, M. and Marsman, M. and Kresse, G.},
	doi = {10.1103/PhysRevLett.99.246403},
	issue = {24},
	journal = {Phys. Rev. Lett.},
	month = {Dec},
	numpages = {4},
	pages = {246403},
	publisher = {American Physical Society},
	title = {Accurate Quasiparticle Spectra from Self-Consistent {{$GW$}} Calculations with Vertex Corrections},
	url = {https://link.aps.org/doi/10.1103/PhysRevLett.99.246403},
	volume = {99},
	year = {2007}}

@article{DelSol_1994,
	author = {Del Sole, R. and Reining, Lucia and Godby, R. W.},
	doi = {10.1103/PhysRevB.49.8024},
	issue = {12},
	journal = {Phys. Rev. B},
	month = {Mar},
	numpages = {0},
	pages = {8024--8028},
	publisher = {American Physical Society},
	title = {{{$GW\Gamma$}} approximation for electron self-energies in semiconductors and insulators},
	url = {https://link.aps.org/doi/10.1103/PhysRevB.49.8024},
	volume = {49},
	year = {1994}}

@article{Morris_2007,
	author = {Morris, Andrew J. and Stankovski, Martin and Delaney, Kris T. and Rinke, Patrick and Garc\'{\i}a-Gonz\'alez, P. and Godby, R. W.},
	doi = {10.1103/PhysRevB.76.155106},
	issue = {15},
	journal = {Phys. Rev. B},
	month = {Oct},
	numpages = {9},
	pages = {155106},
	publisher = {American Physical Society},
	title = {Vertex corrections in localized and extended systems},
	url = {https://link.aps.org/doi/10.1103/PhysRevB.76.155106},
	volume = {76},
	year = {2007}}

@article{Mejuto-Zaera_2022,
	author = {Mejuto-Zaera, Carlos and Vl\ifmmode \check{c}\else \v{c}\fi{}ek, Vojt\ifmmode \check{e}\else \v{e}\fi{}ch},
	doi = {10.1103/PhysRevB.106.165129},
	issue = {16},
	journal = {Phys. Rev. B},
	month = {Oct},
	numpages = {8},
	pages = {165129},
	publisher = {American Physical Society},
	title = {Self-consistency in {{$GW\Gamma$}} formalism leading to quasiparticle-quasiparticle couplings},
	url = {https://link.aps.org/doi/10.1103/PhysRevB.106.165129},
	volume = {106},
	year = {2022}}

@article{Li_2022,
	author = {Li, Jiachen and Jin, Ye and Rinke, Patrick and Yang, Weitao and Golze, Dorothea},
	doi = {10.1021/acs.jctc.2c00617},
	journal = {J. Chem. Theory Comput.},
	number = {12},
	pages = {7570-7585},
	title = {Benchmark of {{$GW$}} Methods for Core-Level Binding Energies},
	volume = {18},
	year = {2022}}

@article{Tolle_2022,
	author = {Johannes T{\"o}lle and Garnet Kin-Lic Chan},
	doi = {10.1063/5.0139716},
	journal = {J. Chem. Phys.},
	month = {3},
	pages = {124123},
	title = {Exact relationships between the {{$GW$}} approximation and equation-of-motion coupled-cluster theories through the quasi-boson formalism},
	volume = {158},
	year = {2023}}

@article{Forster_2020,
	author = {F{\"o}rster, Arno and Visscher, Lucas},
	doi = {10.1021/acs.jctc.0c00693},
	journal = {J. Chem. Theory Comput.},
	number = {12},
	pages = {7381-7399},
	title = {Low-Order Scaling {{$G_0W_0$}} by Pair Atomic Density Fitting},
	volume = {16},
	year = {2020}}

@article{McKeon_2022,
	author = {McKeon,Caroline A. and Hamed,Samia M. and Bruneval,Fabien and Neaton,Jeffrey B.},
	doi = {10.1063/5.0097582},
	journal = {J. Chem. Phys.},
	number = {7},
	pages = {074103},
	title = {An optimally tuned range-separated hybrid starting point for ab initio {{$GW$}} plus {{Bethe--Salpeter}} equation calculations of molecules},
	volume = {157},
	year = {2022}}

@book{Shavitt_2009,
	address = {{Cambridge}},
	author = {Shavitt, Isaiah and Bartlett, Rodney J.},
	doi = {10.1017/CBO9780511596834},
	isbn = {978-0-521-81832-2},
	publisher = {{Cambridge University Press}},
	series = {Cambridge {{Molecular Science}}},
	title = {Many-{{Body Methods}} in {{Chemistry}} and {{Physics}}: {{MBPT}} and {{Coupled}}-{{Cluster Theory}}},
	year = {2009}}

@article{Bartlett_2007,
	author = {Bartlett, Rodney J. and Musia{\l}, Monika},
	doi = {10.1103/RevModPhys.79.291},
	journal = {Rev. Mod. Phys.},
	pages = {291--352},
	title = {Coupled-Cluster Theory in Quantum Chemistry},
	volume = {79},
	year = {2007}}

@incollection{Crawford_2000,
	author = {Crawford, T. Daniel and Schaefer, Henry F.},
	booktitle = {Reviews in {{Computational Chemistry}}},
	doi = {10.1002/9780470125915.ch2},
	isbn = {978-0-470-12591-5},
	pages = {33--136},
	publisher = {{John Wiley \& Sons, Ltd}},
	title = {An {{Introduction}} to {{Coupled Cluster Theory}} for {{Computational Chemists}}},
	year = {2000}}

@article{Monino_2022,
	author = {Monino,Enzo and Loos,Pierre-Fran{\c c}ois},
	doi = {10.1063/5.0089317},
	journal = {J. Chem. Phys.},
	number = {23},
	pages = {231101},
	title = {Unphysical discontinuities, intruder states and regularization in {{$GW$}} methods},
	volume = {156},
	year = {2022}}

@article{Forster_2021,
	author = {F{\"o}rster, Arno and Visscher, Lucas},
	doi = {10.3389/fchem.2021.736591},
	journal = {Front. Chem.},
	pages = {736591},
	title = {Low-Order Scaling Quasiparticle Self-Consistent {{$GW$}} for Molecules},
	volume = {9},
	year = {2021}}

@article{Golze_2018,
	author = {Golze, Dorothea and Wilhelm, Jan and van Setten, Michiel J. and Rinke, Patrick},
	doi = {10.1021/acs.jctc.8b00458},
	journal = {J. Chem. Theory Comput.},
	number = {9},
	pages = {4856-4869},
	title = {Core-Level Binding Energies from {{$GW$}}: An Efficient Full-Frequency Approach within a Localized Basis},
	volume = {14},
	year = {2018}}

@article{Golze_2020,
	author = {Golze, Dorothea and Keller, Levi and Rinke, Patrick},
	doi = {10.1021/acs.jpclett.9b03423},
	journal = {J. Phys. Chem. Lett.},
	number = {5},
	pages = {1840-1847},
	title = {Accurate Absolute and Relative Core-Level Binding Energies from {{$GW$}}},
	volume = {11},
	year = {2020}}

@article{Duchemin_2021,
	author = {Duchemin, Ivan and Blase, Xavier},
	doi = {10.1021/acs.jctc.1c00101},
	journal = {J. Chem. Theory Comput.},
	number = {4},
	pages = {2383-2393},
	title = {Cubic-Scaling All-Electron {{$GW$}} Calculations with a Separable Density-Fitting Space--Time Approach},
	volume = {17},
	year = {2021}}

@article{Bruneval_2021,
	author = {Bruneval, Fabien and Dattani, Nike and van Setten, Michiel J.},
	doi = {10.3389/fchem.2021.749779},
	journal = {Front. Chem.},
	pages = {749779},
	title = {The {{$GW$}} Miracle in Many-Body Perturbation Theory for the Ionization Potential of Molecules},
	volume = {9},
	year = {2021}}

@article{Bintrim_2021,
	author = {Bintrim,Sylvia J. and Berkelbach,Timothy C.},
	doi = {10.1063/5.0035141},
	journal = {J. Chem. Phys.},
	number = {4},
	pages = {041101},
	title = {Full-frequency {{$GW$}} without frequency},
	volume = {154},
	year = {2021}}

@article{Blase_2020,
	author = {Blase, Xavier and Duchemin, Ivan and Jacquemin, Denis and Loos, Pierre-Fran{\c c}ois},
	doi = {10.1021/acs.jpclett.0c01875},
	journal = {J. Phys. Chem. Lett.},
	number = {17},
	pages = {7371-7382},
	title = {The {Bethe--Salpeter} Equation Formalism: {From} Physics to Chemistry},
	volume = {11},
	year = {2020}}

@article{Liu_2020,
	author = {C. Liu and J. Kloppenburg and Y. Yao and X. Ren and H. Appel and Y. Kanai and V. Blum},
	doi = {10.1063/1.5123290},
	journal = {J. Chem. Phys.},
	pages = {044105},
	title = {All-electron ab initio {{Bethe-Salpeter}} equation approach to neutral excitations in molecules with numeric atom-centered orbitals},
	volume = {152},
	year = {2020}}

@article{Piecuch_2002,
	author = {Piotr Piecuch and Karol Kowalski and Ian S. O. Pimienta and Michael J. Mcguire},
	doi = {10.1080/0144235021000053811},
	journal = {Int. Rev. Phys. Chem.},
	pages = {527-655},
	publisher = {Taylor & Francis},
	title = {Recent advances in electronic structure theory: Method of moments of coupled-cluster equations and renormalized coupled-cluster approaches},
	volume = {21},
	year = {2002}}

@article{Dvorak_2019b,
	author = {M. Dvorak and D. Golze and P. Rinke},
	doi = {10.1103/PhysRevMaterials.3.070801},
	journal = {Phys. Rev. Mat.},
	pages = {070801(R)},
	title = {Quantum embedding theory in the screened Coulomb interaction: Combining configuration interaction with {{$GW$/BSE}}},
	volume = {3},
	year = {2019}}

@article{Dvorak_2019a,
	author = {M. Dvorak and P. Rinke},
	doi = {10.1103/PhysRevB.99.115134},
	journal = {Phys. Rev. B},
	pages = {115134},
	title = {Dynamical configuration interaction: Quantum embedding that combines wave functions and Green's functions},
	volume = {99},
	year = {2019}}

@article{Baumeier_2012a,
	author = {Baumeier, Bj\"{o}rn and Andrienko, Denis and Rohlfing, Michael},
	doi = {10.1021/ct300311x},
	journal = {J. Chem. Theory Comput.},
	number = {8},
	pages = {2790-2795},
	title = {Frenkel and Charge-Transfer Excitations in Donor--Acceptor Complexes From Many-Body {{Green's}} Functions Theory},
	volume = {8},
	year = {2012}}

@misc{QuAcK,
	author = {P. F. Loos},
	doi = {10.5281/zenodo.3745928},
	note = {\url{https://github.com/pfloos/QuAcK}},
	publisher = {Zenodo},
	title = {{{QuAcK: a software for emerging quantum electronic structure methods}}},
	url = {https://github.com/pfloos/QuAcK},
	year = {2019}}

@article{Blase_2011a,
	author = {Blase,X. and Attaccalite,C.},
	doi = {10.1063/1.3655352},
	journal = {Appl. Phys. Lett.},
	number = {17},
	pages = {171909},
	title = {Charge-Transfer Excitations in Molecular Donor-Acceptor Complexes within the Many-Body {{Bethe-Salpeter}} Approach},
	volume = {99},
	year = {2011}}

@article{Blase_2018,
	author = {Blase, Xavier and Duchemin, Ivan and Jacquemin, Denis},
	doi = {10.1039/C7CS00049A},
	issue = {3},
	journal = {Chem. Soc. Rev.},
	pages = {1022-1043},
	publisher = {The Royal Society of Chemistry},
	title = {The {Bethe--Salpeter} equation in chemistry: relations with {TD-DFT}, applications and challenges},
	url = {http://dx.doi.org/10.1039/C7CS00049A},
	volume = {47},
	year = {2018}}

@article{Bruneval_2015,
	author = {Bruneval,Fabien and Hamed,Samia M. and Neaton,Jeffrey B.},
	doi = {10.1063/1.4922489},
	journal = {J. Chem. Phys.},
	number = {24},
	pages = {244101},
	title = {A systematic benchmark of the ab initio {{Bethe-Salpeter}} equation approach for low-lying optical excitations of small organic molecules},
	volume = {142},
	year = {2015}}

@article{Bruneval_2016,
	author = {Bruneval, Fabien and Rangel, Tonatiuh and Hamed, Samia M. and Shao, Meiyue and Yang, Chao and Neaton, Jeffrey B.},
	doi = {10.1016/j.cpc.2016.06.019},
	issn = {00104655},
	journal = {Comput. Phys. Commun.},
	month = nov,
	pages = {149--161},
	shorttitle = {Molgw 1},
	title = {Molgw 1: {{Many}}-Body Perturbation Theory Software for Atoms, Molecules, and Clusters},
	volume = {208},
	year = {2016}}

@article{Caruso_2016,
	author = {F. Caruso and M. Dauth and M. J. {van Setten} and P. Rinke},
	doi = {10.1021/acs.jctc.6b00774},
	journal = {J. Chem. Theory Comput.},
	pages = {5076},
	title = {Benchmark of {{$GW$}} Approaches for the {{$GW$}}100 Test Set},
	volume = {12},
	year = {2016}}

@article{DiSabatino_2016,
	author = {Di Sabatino, Stefano and Berger, J. A. and Reining, Lucia and Romaniello, Pina},
	doi = {10.1103/PhysRevB.94.155141},
	issn = {2469-9950, 2469-9969},
	journal = {Physical Review B},
	month = oct,
	number = {15},
	pages = {155141},
	shorttitle = {Photoemission Spectra from Reduced Density Matrices},
	title = {Photoemission Spectra from Reduced Density Matrices: {{The}} Band Gap in Strongly Correlated Systems},
	volume = {94},
	year = {2016}}

@article{Faber_2011,
	author = {Faber, Carina and Attaccalite, Claudio and Olevano, V. and Runge, E. and Blase, X.},
	doi = {10.1103/PhysRevB.83.115123},
	issn = {1098-0121, 1550-235X},
	journal = {Phys. Rev. B},
	month = mar,
	number = {11},
	pages = {115123},
	title = {First-Principles {{$GW$}} Calculations for {{DNA}} and {{RNA}} Nucleobases},
	volume = {83},
	year = {2011}}

@article{Godby_1986,
	author = {Godby, R. W. and Schl{\"u}ter, M. and Sham, L. J.},
	doi = {10.1103/PhysRevLett.56.2415},
	journal = {Phys. Rev. Lett.},
	number = {22},
	pages = {2415--2418},
	title = {Accurate {{Exchange-Correlation Potential}} for {{Silicon}} and {{Its Discontinuity}} on {{Addition}} of an {{Electron}}},
	volume = {56},
	year = {1986}}

@article{Godby_1987a,
	author = {Godby, R. W. and Schl{\"u}ter, M. and Sham, L. J.},
	doi = {10.1103/PhysRevB.35.4170},
	journal = {Phys. Rev. B},
	number = {8},
	pages = {4170--4171},
	title = {Quasiparticle Energies in {{GaAs}} and {{AlAs}}},
	volume = {35},
	year = {1987}}

@article{Jacquemin_2017a,
	author = {Jacquemin, Denis and Duchemin, Ivan and Blase, Xavier},
	doi = {10.1021/acs.jpclett.7b00381},
	journal = {J. Phys. Chem. Lett.},
	number = {7},
	pages = {1524-1529},
	title = {{{Is the Bethe--Salpeter Formalism Accurate for Excitation Energies? Comparisons with TD-DFT, CASPT2, and EOM-CCSD}}},
	volume = {8},
	year = {2017}}

\end{document}